\documentclass[%
 reprint,
 amsmath,amssymb,
 aps,
]{revtex4-2}

\usepackage{mathtools}
\usepackage{graphicx}
\usepackage{dcolumn}
\usepackage{bm}

\usepackage{color}
\usepackage{comment}

\newcommand{\bra}[1]{\langle#1\rvert} 
\newcommand{\ket}[1]{\lvert#1\rangle} 
\newcommand{\qprod}[2]{ \langle #1 | #2 \rangle} 

\usepackage[normalem]{ulem}

\begin{document}


\title{Universality of critical dynamics on a complex network}

\author{Mrinal Sarkar$^1$}
\email[\\Corresponding author: ]{sarkar@thphys.uni-heidelberg.de}
\author{Tilman Enss$^1$}%
\author{Nicolò Defenu$^2$}%
\affiliation{
$^1$Institut für Theoretische Physik, Universität Heidelberg, Philosophenweg 19, 69120 Heidelberg, Germany 
}%
\affiliation{
$^2$Institut für Theoretische Physik, ETH Zürich, Wolfgang-Pauli-Str.27, 8093 Zürich, Switzerland
}%

\date{\today}

\begin{abstract}
We investigate the role of the spectral dimension $d_s$ in determining the universality of phase transitions on a complex network. Due to its structural heterogeneity, a complex network generally acts as a disordered system. Specifically, we study the synchronization and entrainment transitions in the nonequilibrium dynamics of the Kuramoto model and the phase transition of the equilibrium dynamics of the classical $XY$ model, thereby covering a broad spectrum from nonlinear dynamics to statistical and condensed matter physics. Using linear theory, we obtain a general relationship between the dynamics occurring on the network and the underlying network properties. This yields the lower critical spectral dimension of the phase synchronization and entrainment transitions in the Kuramoto model as $d_s=4$ and $d_s=2$ respectively, whereas for the phase transition in the $XY$ model it is $d_s=2$. To test our theoretical hypotheses, we employ a network where any two nodes on the network are connected with a probability proportional to a power law of the distance between the nodes; this realizes any desired $d_s\in [1, \infty)$. Our detailed numerical study agrees well with the prediction of linear theory for the phase synchronization transition in the Kuramoto model. However, it shows a clear entrainment transition in the Kuramoto model and phase transition in the $XY$ model at $d_s \gtrsim 3$, not $d_s=2$ as predicted by linear theory. Our study indicates that network disorder in the region $2 \leq d_s \lesssim 3$ introduces strong finite-size fluctuations, which makes it extremely difficult to probe the existence of the ordered phase as predicted, affecting the dynamics profoundly.
\end{abstract}

\maketitle

\section{Introduction}
\label{sec:intro}
Phase transitions and critical phenomena are a much-studied research topic in statistical and condensed matter physics. By now, it is well known how large-scale geometry affects the universal behavior of phase transitions and critical phenomena. Most of the research, both in and out of equilibrium, focuses on regular Euclidean lattices. Renormalization group theory helps us understand this phenomenon for statistical systems on such a regular lattice. It predicts that the Euclidean dimension is the only relevant geometrical parameter in determining universality classes. 

However, the situation is not so obvious when we have a system that breaks the translational invariance, such as disordered lattices, fractals, amorphous materials, or, in general, graphs or networks. Although a vast amount of literature has been dedicated to studying critical phenomena on disordered lattices in condensed matter and statistical physics, the same on a general graph or a  complex network appears new to this list, drawing attention only recently. A complex network, formed by a set of nodes and links between them, also acts as a disordered system, where the disorder arises from structural heterogeneity: the degree distribution and its various moments~\cite{boccaletti2006complex,networks_newman}. Based on the structural properties or inhomogeneity of the network, the critical dynamics on such a network may yield nontrivial and intriguing results.

Critical dynamics on complex networks finds many applications in diverse fields: epidemic spreading~\cite{apolloni2014metapopulation}, brain dynamics~\cite{bullmore2009complex}, urban traffic on roads~\cite{petri2013entangled}, and many others~\cite{dorogovtsev2008critical}. A pertinent question is: Which parameter determines the universality classes for critical systems on a general network? It is believed that the spectral dimension, more specifically, the ``average spectral dimension" ($d_s$) of a network, is equivalent to the Euclidean dimension for lattices. The spectral dimension characterizes the scaling of the low-lying eigenvalues of the associated Laplacian of the network~\cite{van2023graph, latora2017complex}. If the first, smallest non-zero (Fiedler) eigenvalue vanishes in the thermodynamic limit, the network is said to have a finite spectral dimension~\cite{dhar1977lattices,burioni1996universal}. On the other hand, if it remains finite, the network is said to develop a spectral gap, implying infinite spectral dimension.

We ask a fundamental question: Given two networks with the same spectral dimension $d_s$, does the dynamics on both networks belong to the same universality class? Or, in other words, is the spectral dimension the only relevant parameter of a network that determines the universality class of the dynamics? This quest has been pursued in the field of statistical physics for a long time; however, a unique answer to it is still lacking. The relation between network geometry and dynamics is far from trivial. There are a few studies that generalize the analysis of lattices to networks. We must note that a complex network is fundamentally very different from a regular Euclidean lattice. For a network and a lattice with the same finite $d_s$, the eigenvectors of the network Laplacian may not be delocalized over the network like the Fourier basis on the Euclidean lattice. For a clean system (a lattice), all eigenvectors are delocalized; however, for a disordered graph (a complex network) of finite size, some states may appear localized, reflecting local configurations or degree distributions. This raises the fundamental question about the role of spectral dimension in determining the universality of the dynamics. Addressing this question is the key focus of our present work. 

The nonlinearity in the dynamics and the heterogeneity in a complex network make, in general, such a study analytically formidable~\cite{dorogovtsev2008critical}. Even numerically, one major problem lies in accessing networks of any spectral dimensions. To overcome this, we have employed a network where we can tune its spectral dimension continuously and thus have the freedom to work in any dimension $d_s \in [1, \infty)$ and test numerically various hypotheses or theoretical predictions~\cite{millan2021complex}.

To study the role of spectral dimension in determining the universality of the dynamics, we consider two fundamentally different kinds of dynamics on such a network: One is a nonequilibrium dynamics of a paradigmatic model of nonlinear dynamics, namely, the Kuramoto model showing spontaneous synchronization~\cite{Kuramotobook,Strogatz2000,Pikovskybook,kuramotormp}, and the other one is an equilibrium one, namely, the dynamics of a classical $XY$ model, another paradigmatic model of statistical and condensed matter physics. First, we theoretically obtain the role of $d_s$ on the phase and entrainment transition for the Kuramoto dynamics and test the theoretical predictions numerically. Next, we investigate the role of $d_s$ for the universality of the phase transition of the classical $XY$ model~\cite{cassi1992phase,cassi1996local}. 
One advantage of studying the Kuramoto model is that we can map it to the dynamics of the classical $XY$ model in a certain limit: Making the natural frequencies of the oscillators identical, then going to a co-moving frame rotating at that identical frequency, and applying Gaussian white noise to the systems~\cite{sarkar2020noise,sarkar2021synchronization}. This is equivalent to making the `quenched' disorder of the natural frequencies an `annealed' one (Gaussian white noise). In this paper, we examine both cases on networks in turn; more details will be provided in the subsequent section.

We show in this work for a given dynamics occurring on the network, under linear approximation, how the stationary-state fluctuations and phase correlations depend on the underlying network properties, namely, the density of eigenvalues of the network Laplacian and the stationary distribution of the coefficients (corresponding to the variables expressed in the eigenbasis of the Laplacian). In the context of the Kuramoto model and the classical $XY$ model, we explicitly obtain these quantities, which further help us estimate the lower critical spectral dimension $d_s$ of the associated phase transitions. Our main finding is that the linear theory predicts the lower critical dimension for entrainment and synchronization transition in the Kuramoto model as $d_s=2$, and $d_s=4$, respectively,  and for the phase transition in the $XY$ model it is $d_s=2$. Our detailed numerical investigation agrees well with the theoretical prediction of synchronization transition in the Kuramoto model. However, it does not yield a clear signature of entrainment transition in the Kuramoto model and phase transition in the $XY$ model in $2 \leq d_s \lesssim 3$. This indicates that the network heterogeneity in the form of bond disorder is harmless at dimension $d_s > 3$, whereas it plays a crucial role in the region $2 \leq d_s \lesssim 3$: It introduces strong finite-size fluctuations, which vanishes very slowly in the thermodynamic limit. This, in turn, makes it extremely difficult to probe the existence of the ordered phase as predicted\, \cite{vezzani1999inverse}.

For a similar line of work, we cite Ref.~\cite{millan2019synchronization} in the context of synchronization in the Kuramoto model at various spectral dimensions. Our theoretical results for the Kuramoto model match the prediction in Ref.~\cite{millan2019synchronization}, which validates our theory. However, our work is different from that in three contexts: First, we go beyond the Ref.~\cite{millan2019synchronization} in that we have derived the expressions of the observables on a general ground, under the assumption that the dynamics (of interest) has a unique stationary state on the network, while it was obtained in the reference using an explicit solution of the deterministic Kuramoto dynamics. Second, our method is general in that it applies to both deterministic and stochastic dynamics on the network so long as it has a unique stationary state. Third, in Ref.~\cite{millan2019synchronization}, the numerical test was performed on a complex network manifold that generates discrete $d$-dimensional manifolds by gluing $d$-dimensional simplices along their $(d-1)$-faces subsequently, whereas, in numerics, we tune spectral dimension in our model continuously. Another work in the similar line but for the second-order Kuramoto model on a lattice can be found in Ref.~\cite{odor2023synchronization}.

In passing, we note that the heterogeneous degree distribution makes the Kuramoto model under study a `disordered' system, specifically in terms of bond disorder. This is very different from other `disordered' Kuramoto models studied in the literature, where the disorder arises from independent random positive and negative mean-field couplings, which add frustration to the system and can lead to glassy dynamics\,\cite{daido1992quasientrainment, bonilla1993glassy,pruser2024nature}.

The paper is organized as follows. Sec~\ref{sec:model} describes the graph/network we work on and defines our model of study along with the main queries addressed in this work. In Sec~\ref{sec:linear_th}, we derive our theoretical predictions for the stability of the ordered phase in terms of spectral dimension on a general network under linear approximation, thereby predicting the lower critical dimensions for the phase and entrainment transitions in the Kuramoto and XY models. Sec~\ref{sec:observables} defines the observables to study numerically the associated transitions of the Kuramoto and XY dynamics. In Sec~\ref{sec:Kura}, we provide our numerical results on phase and entrainment transitions of the Kuramoto dynamics, test our theoretical predictions, and discuss the possible role of graph disorder on the dynamics, while in Sec~\ref{sec:xy} we study the same for the XY dynamics. The paper ends with conclusions in Sec.~\ref{sec:conclusion}. Finally, Appendix~\ref{app:Linear_theory} provides a derivation of the quantities required to compute fluctuations and correlation in the linear theory.

\section{Model and dynamics}
\label{sec:model}

In this section, we first introduce the model, namely, the 1D Long-range random ring (1DLRRR or 1DLR3) network~\cite{millan2021complex}. The network is constructed as follows: We first consider as a backbone a one-dimensional linear lattice of $N$ sites with periodic boundary conditions, thus forming a ring. Next, any two sites $i$ and $j~(\neq i)$ ($i,j = 0,1, 2, \cdots, N-1$) on the lattice are connected by a link with a probability $p_{ij} = 1/|i-j|^{1+\sigma}$, where $|i-j| = \min(|i-j|, N-|i-j|)$
, and the parameter $\sigma$ characterizes the scaling of the  probability $p_{ij}$ with distance. The network does not contain any self-loops. The network is characterized by the adjacency matrix $A= \{ a_{ij}\}$, with $ a_{ij} = 1$ or $0$ depending on whether the nodes $i$ and $j$ are connected or not, according to the link probability. Furthermore, we consider the network to be undirected and symmetric, i.e., $a_{ji} = a_{ij}$. The parameter $\sigma$ controls the network properties: in particular, the degree distribution. By tuning the parameter $\sigma$, one could generate a sparse or dense network, or a network of tightly connected local networks with rare long-range links. For example, $\sigma = -1$ corresponds to a network with \textit{all-to-all} connections, whereas $\sigma \to \infty$ corresponds to a $1$D lattice with nearest neighbor connections only, and one can thus obtain a network with local and long-range links by tuning $\sigma$ between these two extreme limits. We refer the reader to Ref.~\cite{millan2021complex} for the basic network characteristics, e.g., the degree distribution and its mean, variance, etc., of this model.

In contrast to the well-studied 1D lattice model with long-range power-law decaying interactions~\cite{gupta2017world}, the present model is a network (1D lattice with additional long-range links) where the interaction strength is equal among all neighbors; however, the connection probability decays as a power-law. The coupling probability gives rise to randomness that violates the translational invariance and also acts as a source of quenched `bond-disorder' in the system. This model has already been employed to investigate the critical properties of long-range epidemics~\cite{grassberger2013sir} and percolation~\cite{gori2017one}. Also, critical dynamics of the XY model~\cite{berganza2013critical,cescatti2019analysis}, epidemic spreading~\cite{grassberger2013two} and self-avoiding walks~\cite{bighin2024universal} have been studied on a two-dimensional lattice version of it. A generalization of the present model is reminiscent of the Kleinberg model~\cite{kleinberg2000navigation} of network science, showing the emergence of the small-world phenomenon~\cite{watts1998collective}, and was employed in the study of navigation problems~\cite{carmi2009asymptotic, cartozo2009extended, li2010towards}.

The spectral properties of our 1DLR3 network are thoroughly investigated in Ref.~\cite{millan2021complex}. By tuning the parameter $\sigma$ continuously, this model allows one to realize the whole range of spectral dimension $d_s \in [1, ~\infty)$. This feature makes it a suitable candidate to investigate the universal behavior for critical models with both continuous and discrete symmetries.

To study the universality of the dynamics on such a network, we first work with the paradigmatic model of nonlinear dynamics, namely, the Kuramoto model, showing the spontaneous emergence of collective synchronization~\cite{Kuramotobook,Strogatz2000,Pikovskybook,kuramotormp}. The model comprises a collection of $N$ interacting limit-cycle oscillators residing at nodes of the network and of distributed natural frequencies. The phase $\theta_i(t) \in [0,2\pi)$ of the $i$-th oscillator evolves in time as~\cite{Kuramotobook} 
\begin{align}
	\frac{{\rm d} \theta_{i}}{{\rm d}t}=\omega_{i}+\frac{K}{\kappa_i}\sum_{j=1}^N  a_{ij} \sin(\theta_{j} -\theta_{i}).
	\label{eq:eom0}
\end{align}
Here, $K \ge 0$ denotes the strength of coupling between the oscillators,  $\omega_i \in (-\infty,\infty)$ is the natural frequency of the $i$-th oscillator,  $a_{ij}$ is the adjacency matrix for a given network realization, and $\kappa_i=\sum_j a_{ij}$ is the degree of the $i$-th node. The scaling by $\kappa_i$ of the second term on the right-hand side ensures that this term is well behaved in the limit $N \to \infty$ and, moreover, to screen out the effect of having heterogeneous degree distributions. The $\omega_{j}$'s are quenched-disordered random variables distributed according to a common distribution $G(\omega)$ with finite mean $\Omega_0$ and width $\Delta>0$. By choice of a suitable frame of reference, the mean of the distribution $\Omega_0$ can be set to zero without loss of generality. In our numerical simulations, the natural frequencies are drawn from a Gaussian distribution with zero mean and unit variance.

Note that the dynamics~(\ref{eq:eom0}) is deterministic and is, moreover, intrinsically non-Hamiltonian. The latter fact is due to the presence of two sources of quenched disorder: the degree ($\kappa_i$) and the intrinsic frequency ($\omega_i$). Even without the frequency term, the asymmetric term $a_{ij}/\kappa_i$ in the interaction kernel cannot be derived from a potential flow, rendering the dynamics non-Hamiltonian. Consequently, the dynamics~(\ref{eq:eom0}) always relaxes at long times to a nonequilibrium stationary state.

Let us now briefly summarize the known results for the stationary state of the dynamics~(\ref{eq:eom0}) in various limits. For $\sigma \leq 0$, the 1DLR3 network develops a finite spectral gap indicating spectral dimension $d_s = \infty$~\cite{millan2021complex,burioni1997geometrical}. For $\sigma = -1$, $a_{ij}=1~\forall i,j$, the dynamics reduces to that of a mean-field, \textit{all-to-all} coupled model, originally introduced by Kuramoto \cite{Kuramotobook}. Depending on the value of the coupling $K$, this model exhibits two qualitatively different phases in the thermodynamic limit: a low-$K$ unsynchronized phase where the oscillators run incoherently (a more precise definition will be given in the following section) and a high-$K$ synchronized phase where a macroscopic number of oscillators or even all of them lock their frequencies despite having different natural frequencies, and run coherently. The dynamics (\ref{eq:eom0}) exhibits a supercritical bifurcation between these two phases as one tunes $K$ across a critical value $K_{c}$~\cite{Kuramotobook, Strogatz2000, Gupta2018}. By analogy with a statistical system, we may associate the bifurcation behavior with a continuous phase transition~\cite{Gupta2018, Livi2017}. From now on, we use ``phase transition'' instead of ``bifurcation'' throughout this paper. One would expect a similar phase transition for $\sigma \leq 0$. On the other hand, in the opposite limit $\sigma \to \infty$, the dynamics is equivalent to that on a one-dimensional chain with a nearest-neighbor interaction, which in the limit $N \to \infty$ does not exhibit any ordered phase at any $K$ and hence no phase transition~\cite{Strogatz2000}.

For $\sigma >0$, the model has a finite spectral dimension: based on the study in Ref.~\cite{millan2021complex}, one expects $d_s = 1$ for $\sigma \geq 2$ and $d_s = 2/ \sigma$ for $0 < \sigma < 1/3$, the same as on a fully connected weighted graph~\cite{burioni1997geometrical}. However, in the range $1/3 \leq \sigma < 2$, the behavior deviates from that of a fully connected weighted graph, and one needs to study the low-energy spectrum to determine $d_s$. In this context, we ask: What is the role of $d_s$ in the synchronization dynamics of the Kuramoto model? How does the network disorder affect the dynamics? Or is the spectral dimension the only relevant parameter determining the universality of the dynamics on such a disordered graph? A thorough investigation to address these questions is one of the primary goals of the present work. A recent study in this direction can be found in Ref.~\cite{millan2019synchronization}, where the synchronization dynamics of the Kuramoto model was studied on a complex network manifold, which is different from ours.

As we will also investigate the role of $d_s$ for the universality of the phase transition of the classical $XY$ model, let us now summarize the known results for the $XY$ model. Similar to the Kuramoto model, it does not exhibit any phase transition in the limit $N \to \infty$ on a 1D lattice with nearest-neighbor interactions, which corresponds to $\sigma \geq 2$ in our model. Note that on a two-dimensional regular lattice, this model undergoes the Berezinskii-Kosterlitz-Thouless  (BKT) phase transition~\cite{kosterlitz1973ordering,kosterlitz1974critical}. Following the Mermin-Wagner theorem, the lower critical dimension for a phase transition is $d^{l}=2$~\cite{mermin1966absence}. A generalization of the Mermin-Wagner theorem for graphs states that spontaneous breaking of continuous symmetry is not possible on a graph that is ``recursive on average'', i.e., on a graph with ``average spectral dimension'' $d_s \leq 2$. Instead, it is possible only on a graph that is ``transient on average'' ($d_s > 2$)~\cite{cassi1992phase,cassi1996local}. The $XY$ model on the Watts-Strogatz small-world network exhibits a mean-field type continuous phase transition. 
A study of the $XY$ model on complex networks with an annealed network approximation shows the existence of a continuous phase transition with the critical temperature being proportional to the second moment of the degree distribution. This implies that the critical temperature is finite only if the second moment is finite~\cite{dorogovtsev2008critical}. 

There have been several recent works analyzing the fate of a $BKT$ quasi-long range ordered phase in a long-range interacting system. The $XY$ dynamics on a 2D long-range interacting systems, which could be thought of as an `annealed' version of our model in 2D, where the interaction decays as $\sim |i-j|^{-(2+\sigma)}$, yields a rich phase diagram showing the existence of both conventional continuous and $BKT$ transitions in the region $7/4 < \sigma < 2$, and belonging to the $BKT$ universality class for $\sigma >2$~\cite{giachetti2021berezinskii}. Another recent study of the $XY$ dynamics on a two-dimensional version of a variant of our model, where the connection probability $\sim |i-j|^{-(2+\sigma)}$, verifies the theoretical prediction that the dynamics belongs to the BKT universality class for $\sigma \geq 2$ and exhibits a continuous phase transition for $\sigma < 2$~\cite{berganza2013critical,cescatti2019analysis}. Also, a $BKT$ transition is predicted in the critical dynamics of the $XY$ model on a 1D long-range power-law interacting system for $d_s=2$~\cite{kosterlitz1976phase, brown1988monte}.

The fact that the underlying topology of the network plays a crucial role in the critical dynamics and the advantage that one can realize spectral dimensions $d_s$ lower than $2$ in our 1DLR3 model motivate us to investigate the critical dynamics of the model on a network with $d_s=2$. Investigating $BKT$ on such a graph with $1D$ backbone would be a good test of the universality. Further, it would be interesting to see how the topological excitations, in case a quasi-ordered phase exists, are formed on such a network. This constitutes the second part of our work, where we explore the dynamics of the $XY$ model on the 1DLR3 graph and test the theoretical prediction for the dependence of the universal behavior of BKT and conventional phase transitions on the spectral dimension.

\section{Linear theory}
\label{sec:linear_th}
In this section we first study the dynamics under linear approximation on a general network: $\sin(\theta_{j} -\theta_{i}) \approx (\theta_{j} -\theta_{i}),~\forall i, j$. This corresponds to the case of a very strong $K$ value for the Kuramoto model, and very low temperature for the $XY$ model (see Sec.~\ref{sec:xy}). The linearized equation, as obtained from Eq.~(\ref{eq:eom0}), now reads as, 
\begin{align}
	\frac{{\rm d} \theta_{i}}{{\rm d}t}=\omega_{i} - {K} \sum_{j=1}^{N}  {\mathcal L}_{ij} \theta_{j}.
	\label{eq:eom_linear}
\end{align}
Here ${\mathcal L}_{ij}$, defined as
\begin{align}
{\mathcal L}_{ij} \coloneqq \delta_{ij} - \frac{a_{ij}}{\kappa_i},~~i,j = 1, 2, 3, \cdots, N,
\label{eq:Lap_def_asym}
 \end{align}
 is the $(i,j )$-th element of the associated network Laplacian $ \mathbf {\mathcal L}$~~\cite{van2023graph, godsil2001algebraic}. Note that, by definition~(\ref{eq:Lap_def_asym}), the Laplacian $ \mathbf {\mathcal L}$ is asymmetric; however, it can be shown easily that the eigenvalues $\{ \lambda_i\}_{i=1,2,3, \cdots, N}$ of $\mathbf {\mathcal L}$ are real and non-negative, with the smallest one being $\lambda_1 = 0$~\cite{millan2019synchronization}.
 
To analyze the dynamics~(\ref{eq:eom_linear}), we work in the eigenbasis of the asymmetric Laplacian $ {\mathbf {\mathcal L}}$.
If $ \ket {v_{m}^{R}}$ and $\bra {v_{m}^{L}}$ be the right and left eigenvectors corresponding to an eigenvalue $\lambda_m$, we can represent a state given by the phases of the oscillators, $\ket \theta = (\theta_1, \theta_2, \cdots, \theta_{N})^{\intercal}$, in an eigenbasis as follows:
\begin{align}
\ket {\theta }= \sum_{m =1}^{N} \qprod{v_{m}^{L}}{ \theta} \ket{v_{m}^{R}} = \sum_{m=1}^{N} \theta_{\lambda_{m}}^{R}  \ket{v_{m}^{R}},\\
\bra {\theta }= \sum_{m =1}^{N} \qprod{ \theta} {v_{m}^{R}}  \bra{v_{m}^{L}} = \sum_{m=1}^{N} \theta_{\lambda_{m}}^{L}  \bra{v_{m}^{L}},
\label{eq:theta_eigen}
\end{align}
where $x_{\lambda_{m}}^{R} \coloneqq \qprod{v_{m}^{L}}{ x}$, and  $x_{\lambda_{m}}^{L}\coloneqq \qprod{ x} {v_{m}^{R}} $.
Similarly, a given realization of the natural frequencies ($\{ \omega_i\}$) can also be represented by
\begin{align}
\ket {\omega} = \sum_{m=1}^{N} \omega_{\lambda_{m}}^{R}  \ket{v_{m}^{R}},~
\bra {\omega } = \sum_{m=1}^{N} \omega_{\lambda_{m}}^{L}  \bra{v_{m}^{L}}.
\label{eq:omega_eta_eigen}
 \end{align}
Note that the Laplacian $ {\mathbf {\mathcal L}}$ is now diagonalizable by the modal matrix $\mathbf {P}$ as follows:
\begin{equation}
\mathbf {P}^{-1}  {\mathbf {\mathcal L}}  \mathbf {P} = \mathbf {D},
\end{equation}
where $\mathbf {D}$ is a diagonal matrix with elements being the eigenvalues of the Laplacian $0=\lambda_1 < \lambda_2  \leq \lambda_3, \cdots, \leq \lambda_N$. By construction, the right eigenvectors $ \ket {v_{m}^{R}}$ form the columns of  $\mathbf {P}$, whereas the left eigenvectors $ \bra {v_{m}^{L}}$ form the rows of  $\mathbf {P}^{-1}$. These two sets of eigenvectors form a complete basis, are dual to each other, and can be normalized as $\qprod{v_{m}^{L}}{ v_{m'}^{R}} = \delta_{m,m'}$. Moreover, normalization of the eigenvectors guarantees that $\mathbf {P}^{-1} \mathbf {P}  = \mathbf {P} \mathbf {P}^{-1} = \mathbf{I}$.

\subsection{Observables: Phase fluctuations and phase correlations}
We compute two observables, namely, the average fluctuation of the phases and the phase correlation over the entire network, as proposed in Refs.~\cite{hong2005collective, millan2019synchronization}, to characterize the stability of the synchronized/ordered phase. The phase fluctuation is defined by
\begin{align}
W^2 = \frac{1}{N}    \left \langle  \sum_{i=1}^{N} \left[  \theta_i - \overline {\theta} \right]^2  \right \rangle =  \left \langle  \overline {\theta^2} - {\overline {\theta}}^{2} \right \rangle .
\label{eq:W_sq_def}
\end{align}
where $\overline {\theta} \coloneqq (1/N) \sum_{i=1}^{N}  {\theta_i}$, $\overline {\theta^2} \coloneqq (1/N) \sum_{i=1}^{N}  {\theta_{i}^{2}}$ are the spatial averages, and $\langle \cdot \rangle$ denotes the average over realizations. We now express these spatial averages in the eigenbasis coefficients $\theta_{\lambda}^{L,R}$.

We denote $\theta_i = \qprod{i}{\theta}$, where $\ket{i}= (0, \cdots,0, i, 0,\cdots,0)_{N}^{\intercal}$. We thus have
\begin{align}
\overline {\theta} &= \frac{1}{N}\sum_{i=1}^{N}  {\theta_i} = \frac{1}{N} \sum_{i=1}^{N}   \qprod{i}{\theta} 
=\frac{1}{N} \sum_{i=1}^{N}   \sum_{m=1}^{N}  \theta_{\lambda_{m}}^{R} \qprod{i}{v_{m}^{R}} \nonumber\\
&= \frac{1}{N} \theta_{\lambda_{1}}^{R}  \sum_{i=1}^{N}  \qprod{i}{v_1^R} + \frac{1}{N}  \sum_{m=2}^{N} \theta_{\lambda_{m}}^{R}  \sum_{i=1}^{N}  \qprod{i}{v_{m}^{R}}
\label{eq:av_theta_R}
\end{align}

Similarly, we can express the average phase in the left eigenbasis and obtain
\begin{align}
\overline {\theta} & = \frac{1}{N} \sum_{i=1}^{N}   \qprod{\theta}{i} =\frac{1}{N}  \sum_{m=1}^{N} \theta_{\lambda_{m}}^{L}  \sum_{i=1}^{N}  \qprod{v_{m}^{L}}{i}\nonumber\\
&= \frac{1}{N} \theta_{\lambda_{1}}^{L}  \sum_{i=1}^{N}  \qprod{v_1^L}{i} + \frac{1}{N}  \sum_{m=2}^{N} \theta_{\lambda_{m}}^{L}  \sum_{i=1}^{N}  \qprod{v_{m}^{L}}{i}.
\label{eq:av_theta_L}
\end{align}

Next we will derive a few properties of the Laplacian $ {\mathbf {\mathcal L}}$  to simplify the above averages. The fact that $\sum_{j=1}^{N} {\mathbf {\mathcal L}}\ket{j} = 0$, implies
\begin{align}
 \ket{v_1^R} = \sum_{i=1}^{N} \ket{i},~~{\rm corresponding ~to~} \lambda_1=0.
 \label{eq:v1_R}
\end{align}
Consequently,
\begin{align}
 \sum_{i=1}^{N} \qprod{v_m^L}{i}= 0, ~~\forall m \neq 1. 
\label{eq:vm_L}
\end{align}

This result can also be understood from elementary theory of random walks~\cite{van1992stochastic} on a network, where the Laplacian $ {\mathbf {\mathcal L}}^{\intercal}$ of Eq.~(\ref{eq:Lap_def_asym}) is the master operator or generator of the walk on the network. 

Assuming this process to be ergodic, it has a unique stationary state which is given by $\ket{v_{1}^{R}}$ corresponding to $\lambda_1=0$. One can also show that the $ \bra{v_m^L}$ and $\ket{v_m^R}$ of $ {\mathbf {\mathcal L}}$ are related through
\begin{align}
\ket{v_m^L} = \mathbf {\mathcal{K}} \ket{v_m^R} .
\label{eq:LR_ev_rltn}
\end{align}
where $\mathbf {\mathcal{K}} = \text{diag.} (\kappa_1, \cdots, \kappa_N)$. It readily follows from Eqs.~(\ref{eq:v1_R}, \ref{eq:LR_ev_rltn}) that 
\begin{align}
\bra{v_1^L} = \frac{1}{\langle \kappa \rangle N} \sum_{i=1}^{N} \bra{i} \kappa_i,
 \label{eq:v1_L}
\end{align}
so that the orthogonality condition of the eigenvectors is satisfied. Here, $\langle \kappa \rangle = (1/N) \sum_{i=1}^{N} \kappa_i$. Using Eqs.~(\ref{eq:v1_R}, \ref{eq:vm_L}, \ref{eq:v1_L}), one obtains from Eqs.~(\ref{eq:av_theta_R}, \ref{eq:av_theta_L}), respectively,

\begin{align}
\overline {\theta} = \theta_{\lambda_{1}}^{R} + \frac{1}{N}  \sum_{m=2}^{N} \theta_{\lambda_{m}}^{R}  \sum_{i=1}^{N}  \qprod{i}{v_{m}^{R}}~~{\rm and}~~\overline {\theta} = \frac{1}{N} \theta_{\lambda_{1}}^{L}.
\label{eq:av_theta_R_L}
\end{align}

When averaged over ensembles, we have
\begin{align}
 \left \langle  {\overline {\theta}}^{2} \right \rangle  = \frac{1}{N}   \left \langle   \theta_{\lambda_{1}}^{L} \theta_{\lambda_{1}}^{R}  \right \rangle +  \frac{1}{N^2}  \left \langle  \sum_{m=2}^{N} \theta_{\lambda_{1}}^{L} \theta_{\lambda_{m}}^{R}  \sum_{i=1}^{N}  \qprod{i}{v_{m}^{R}}  \right \rangle.
\label{eq:av_theta_RL}
\end{align}

The second term on the right hand side of Eq.~(\ref{eq:av_theta_RL}) will vanish at long times, which can be understood as follows: For any dynamics of type~(\ref{eq:eom_linear}), where $\omega_i$'s act as stochastic force, be it `quenched' (e.g.,when they act as natural frequencies in the Kuramoto model) or `annealed' (e.g, when they represent Gaussian white noise), the evolution equations for $\theta_{\lambda_{m}}^{L/R}$ become decoupled in the eigenbasis of $\mathbf {\mathcal L}$. The $\theta_{\lambda_{m}}^{L/R}$ can be thought of as velocities of Brownian particles with $\omega_{\lambda_{m}}^{L/R}$ being stochastic force and $K \lambda_m$ plays the role of damping constant; see Appendix \ref{app:Linear_theory}. Thus, at long times i.e. $t \to \infty$, the ensemble average $\left \langle \theta_{\lambda_{m}}^{L/R} \right \rangle$ will decay to zero for each $m$, except for $m=1$ for which $\lambda_1=0$, so long as the ensemble average of the stochastic force is zero, i.e., $\left \langle \omega_{\lambda_{m}}^{L/R} \right \rangle = 0$. Thus, in the limit $t \to \infty$, we have from Eq.~(\ref{eq:av_theta_RL}),

\begin{align}
 \left \langle  {\overline {\theta}}^{2} \right \rangle  = \frac{1}{N}   \left \langle   \theta_{\lambda_{1}}^{L} \theta_{\lambda_{1}}^{R}  \right \rangle.
\label{eq:av_theta_RL2}
\end{align}

Now, the average squared phase, when expressed in eigenbasis,
\begin{align}
\overline {\theta^2} & = \frac{1}{N}  \sum_{i=1}^{N}  {\theta_{i}^{2}} = \frac{1}{N} \qprod{\theta}{\theta}= \frac{1}{N} \sum_{m=1}^{N} \sum_{m'=1}^{N} \theta_{\lambda_{m}}^{L} \theta_{\lambda_{m'}}^{R} \qprod{v_{m}^{L}}{v_{m'}^{R}} \nonumber\\
&= \frac{1}{N}\sum_{m=1}^{N} \theta_{\lambda_{m}}^{L} \theta_{\lambda_{m}}^{R}.
\label{eq:av_theta_sq}
\end{align}

We are interested in the phase fluctuations in the stationary state attained at long times. Thus, Eq.~(\ref{eq:W_sq_def}), on using Eqs.~(\ref{eq:av_theta_RL2}, \ref{eq:av_theta_sq}), yields for the phase fluctuations in the stationary state,
\begin{align}
W^2 =   \left \langle   \overline {\theta^2} - {\overline {\theta}}^{2} \right \rangle  =  \frac{1}{N}  \left \langle \sum_{m=2}^{N} \theta_{\lambda_{m}}^{L} \theta_{\lambda_{m}}^{R}\right \rangle = \frac{1}{N} \sum_{m=2}^{N} \left \langle  \theta_{\lambda_{m}}^{L} \theta_{\lambda_{m}}^{R}\right \rangle.
\label{eq:W_sq_eigen}
\end{align}
In the continuum limit, Eq.~(\ref{eq:W_sq_eigen}) can be expressed as
 \begin{align}
W^2 = \int_{\lambda_2}^{\lambda_{\rm max}} {\rm d}\lambda\, \rho(\lambda)  \left \langle  \theta_{\lambda}^{L} \theta_{\lambda}^{R} \right \rangle ,
\label{eq:W_sq_eigen_cont}
\end{align}
where $\rho(\lambda)$ is the density of eigenvalues of the Laplacian ${\mathbf {\mathcal L}}$, $\lambda_2$ denotes the first nonzero (Fiedler) eigenvalue. The quantity
\begin{align}
\left \langle  \theta_{\lambda}^{L} \theta_{\lambda}^{R} \right \rangle =  \int_{-\infty}^{+\infty} \int_{-\infty}^{+\infty} {\rm d} \theta_{\lambda}^{L} {\rm d} \theta_{\lambda}^{R} ~\theta_{\lambda}^{L} \theta_{\lambda}^{R}~  P_{\rm st} (\theta_{\lambda}^{L}, \theta_{\lambda}^{R} ),
\label{eq:theta_L_R_avg}
\end{align}
where $P_{\rm st} (\theta_{\lambda}^{L}, \theta_{\lambda}^{R} )$ is the stationary joint probability distribution of $\theta_{\lambda}^{L}~{\rm and} ~\theta_{\lambda}^{R}$.

We now compute another important quantity, the phase correlation defined by~\cite{millan2019synchronization}
\begin{align}
C = \frac{1}{N}    \left \langle  \qprod{\theta}{{\mathcal {L}} \theta } \right \rangle ,
\label{eq:corr_def}
\end{align}
where the outer brackets $\langle \cdot \rangle$ denote again the average over realizations. Expressing it in eigenbasis we obtain
\begin{align}
C &=\frac{1}{N}    \left \langle   \sum_{m=1}^{N} \lambda_{m} \qprod{\theta}{v_{m}^{R}} \qprod{v_{m}^{L}}{\theta}  \right \rangle = \frac{1}{N}    \left \langle   \sum_{m=1}^{N} \lambda_{m} \theta_{\lambda_{m}}^{L} \theta_{\lambda_{m}}^{R}  \right \rangle \nonumber\\
& =  \frac{1}{N}    \left \langle   \sum_{m=2}^{N} \lambda_{m} \theta_{\lambda_{m}}^{L} \theta_{\lambda_{m}}^{R}  \right \rangle =  \frac{1}{N}  \sum_{m=2}^{N}  \lambda_{m} \left \langle  \theta_{\lambda_{m}}^{L} \theta_{\lambda_{m}}^{R}  \right \rangle,
\label{eq:corr_def_eigen}
\end{align}
as $\lambda_1=0$. Thus at long times, in the continuum limit,
\begin{align}
C&= \int_{\lambda_2}^{\lambda_{\rm max}} {\rm d}\lambda\, \lambda \rho(\lambda)\, \left \langle  \theta_{\lambda}^{L} \theta_{\lambda}^{R}\right \rangle.
\label{eq:corr_eigen_cont}
\end{align}

Note that the above expressions, given by Eq.~(\ref{eq:W_sq_eigen_cont}, \ref{eq:corr_eigen_cont}), are very general in the sense that they follow immediately from unique stationarity of the dynamics on the network. Thus, it holds for any dynamics occurring on the network so long as it reaches a unique stationary state.

Now, in a general network or disordered system, the density of low-lying eigenvalues $\rho(\lambda)$ of the network Laplacian follows the scaling \cite{burioni1996universal}
 \begin{align}
\rho(\lambda) \sim \lambda^{d_s/2 -1} \quad \text{ for } \lambda \ll 1,
\label{eq:rho_lambda}
\end{align}
where $d_s$ is the \textit{spectral dimension} of the network. Furthermore, the smallest non-zero eigenvalue $\lambda_2$ follows the scaling with the network size $N$ as
 \begin{align}
\lambda_2 \propto N^{-2/d_s}.
\label{eq:lambda2_N}
\end{align}
Equations~(\ref{eq:rho_lambda},\ref{eq:lambda2_N}) would help us express the quantities of interest as a function of the spectral dimension $d_s$.

In the following, we now proceed to compute the quantities $W^2$ and $C$ explicitly for the Kuramoto and $XY$ models, in order to estimate the lower critical dimension of the associated transitions.

\subsection {Theoretical prediction for the Kuramoto model}
In this section, we explicitly compute $W^2$ and $C$ for the Kuramoto model and study their bahavior with system size $N$, which in turn helps to estimate the lower critical dimension of the associated transitions. To start with, we first project the linearized Kuramoto dynamics~(\ref{eq:eom_linear}) along the eigenbasis and obtain evolution equations for $\theta_{\lambda}^{L/R}$. One can then obtain $P_{\rm st} (\theta_{\lambda}^{L}, \theta_{\lambda}^{R} )$ by using the Fokker-Planck formalism, and substitute it in Eq.~(\ref{eq:theta_L_R_avg}) to compute $\left \langle  \theta_{\lambda}^{L} \theta_{\lambda}^{R} \right \rangle $. However, we compute $\left \langle  \theta_{\lambda}^{L} \theta_{\lambda}^{R} \right \rangle $ directly using formal solutions as sketched in Appendix \ref{app:Linear_theory}. 
We have in the stationary state
 \begin{align}
  \left \langle  \theta_{\lambda}^{L}  \theta_{\lambda}^{R} \right \rangle =   \frac{1}{K^2 \lambda^2},
 \label{eq:theta_lambda_corr_Kura}
\end{align}
which is substituted in Eq.~({\ref{eq:W_sq_eigen_cont}}) to arrive at
\begin{align}
W^2 = \int_{\lambda_2}^{\lambda_{\rm max}} {\rm d}\lambda\,  \frac{\rho(\lambda)}{K^2 \lambda^2}.
\label{eq:W_sq_eigen_cont2_Kura}
\end{align}
Note that Eq.~(\ref{eq:W_sq_eigen_cont}) and thus Eq.~(\ref{eq:W_sq_eigen_cont2_Kura}) was obtained in Ref.~\cite{millan2019synchronization} from the explicit solution of the linearized Kuramoto model. However, since we have shown the generality of this equation in the previous section, we directly use it here to obtain the fluctuations.

Further, using Eq.~(\ref{eq:rho_lambda}) and Eq.~(\ref{eq:lambda2_N}), we finally obtain
\begin{equation}
W^2 \sim
\begin{cases}
N^{4/d_s -1},& d_s < 4, \\
\ln N& d_s = 4,   \\
\text{const.}& d_s > 4.
\end{cases}
\label{eq:w_sq_Kura_eigen}
\end{equation} 

A stable synchronized phase requires the average phase fluctuations in the stationary state to be finite in the thermodynamic limit, i.e., $W^2 < \infty$ as $N \to \infty$; otherwise, it becomes thermodynamically unstable. Thus, it immediately follows from Eq.~(\ref{eq:w_sq_Kura_eigen}) that the lower critical dimension for the phase synchronization transition in the Kuramoto model is $d_s = 4$.

On the other hand, using Eq.~(\ref{eq:theta_lambda_corr_Kura}) we obtain from Eq.~(\ref{eq:corr_eigen_cont}) for the stationary-state phase correlations
\begin{align}
C = \int_{\lambda_2}^{\lambda_{\rm max}} {\rm d}\lambda\, \frac{\rho(\lambda)}{K^2 \lambda} ,
\label{eq:corr_eigen_kura}
\end{align}
yielding

\begin{equation}
C \sim 
\begin{cases}
N^{2/d_s -1},&d_s < 2,  \\
\ln N & d_s = 2,  \\
\text{const.} & d_s > 2.
\end{cases}
\label{eq:corr_eigen_kura2}
\end{equation} 

The correlation function $C$ essentially represents the mean-square phase difference between the nearest-neighbor oscillators in the network. Thus,
a divergence in $C$ implies that the average nearest-neighbor phase difference diverges, which contradicts the very assumption of the linear theory. In other words, the linear approximation becomes invalid in this case. 

The correlation $C$ provides useful information about the entrainment dynamics. Physically, an entrained phase is possible so long as the average nearest-neighbor phase difference is small, i.e., the linear theory is valid. Thus, the possibility of an entrained phase arises from the finiteness of $C$.
Based on this, it follows from Eq.~(\ref{eq:corr_eigen_kura2}) that entrainment in the Kuramoto model is possible only if the spectral dimension $d_s> 2$.

\subsection {Theoretical prediction for the $XY$ model}
In this section, we study the system-size behavior of $W^2$ and $C$ for the dynamics of classical $XY$ model; see Sec~\ref{sec:xy} for model definition and detailed dicussion on the $XY$ model. Following a similar approach as for the Kuramoto model, for the linearized dynamics of the $XY$-model projected onto the eigenbasis, one arrives in the stationary state
 \begin{align}
 \left \langle  \theta_{\lambda}^{L}  \theta_{\lambda}^{R} \right \rangle =   \frac{T}{K \lambda}.
 \label{eq:theta_lambda_corr_XY}
\end{align}
A detailed derivation is provided in Appendix~\ref{app:Linear_theory}.

On substituting Eq.~(\ref{eq:theta_lambda_corr_XY}) into Eq.~(\ref{eq:W_sq_eigen_cont}), we obtain the phase fluctuations of the XY model as
 \begin{align}
W^2 = \int_{\lambda_2}^{\lambda_{\rm max}} {\rm d}\lambda\, \rho(\lambda) \frac{T}{K \lambda}.
\label{eq:W_sq_eigen_cont2}
\end{align}
Using Eqs.~(\ref{eq:rho_lambda}) and (\ref{eq:lambda2_N}), one obtains from Eq.~(\ref{eq:W_sq_eigen_cont2}) for the stationary-state fluctuations

\begin{equation}
W^2 \sim
\begin{cases}
N^{2/d_s -1}, & d_s < 2,  \\
\ln N & d_s = 2,  \\
\text{const.} & d_s > 2.
\end{cases}
\label{eq:w_sq_xy_eigen}
\end{equation} 
Note that the functional dependence of $W^2$ for the $XY$ model and $C$ of the Kuramoto model on the eigenvalue spectra of the associated network is the same. Based on the discussion in the previous section, it follows from Eq.~(\ref{eq:w_sq_xy_eigen}) that a thermodynamically stable ordered phase is possible only in $d_s >2$, and the marginal case $d_s=2$ marks the lower critical dimension. 

The correlation in the $XY$ model,
\begin{align}
C = \int_{\lambda_2}^{\lambda_{\rm max}} {\rm d}\lambda\, \rho(\lambda) \frac{T}{K} = \frac{T}{K},
\label{eq:corr_eigen_xy}
\end{align}
is a constant for any $T, K >0$ and $d_s < \infty$.
It implies that the linear theory is always valid. This can be understood physically as follows: Since the $XY$ spins are equivalent to Kuramoto oscillators with identical natural frequencies, the fluctuation in the phase-velocity is always zero, thereby they are inherently always `entrained'. This is why only the phase transition is discussed in the context of the $XY$ model. The behavior of $C$ obtained in Eq.~(\ref{eq:corr_eigen_xy}) is thus consistent with the expected properties of the $XY$ model.

Note that the linear theory never suggests the existence of a phase or entrainment transitions in any $d_s$, for any temperature $T$, or coupling strength $K$. It only states that if an ordered/synchronized state is possible, whether this state would be stable or not. A disordered/unentrained phase emerges and hence, a phase transition actually occurs due to the nonlinearity present in the dynamics. In the following section, we therefore study the nonlinear system numerically and test our theoretical prediction for both the Kuramoto and $XY$ models to understand the role of spectral dimension and network disorder in them.

Further, in numerics we tune the network parameter $\sigma$ to realize various spectral dimensions $d_s$. The relation between $\sigma$ and $d_s$ for our network is nontrivial. It is given by $d_s = 2/ \sigma$ and $d_s=1$ for small $\sigma$($\sigma < 1/3$) and large $\sigma (\sigma > 2)$ respectively, in agreement with that of the fully connected $1D$ weighted graph, where the weight factor between two sites $(i,j)$ is $\sim d_{ij}^{-(1+\sigma)}$ with $d_{ij}$ being the Euclidean shortest distance between the sites $(i,j)$ on the graph. For the intermediate values of $\sigma$, one needs to numerically compute $d_s$ from finite-size scaling of low-lying eigenvalues of the graph Laplacian. The dimensions $d_s=4$ and $2$ correspond to approximately $\sigma=0.5$ and $0.875$, respectively; see Ref.~{\cite{millan2021complex}} and Fig.~{\ref{fig:ds_vs_sigma}}.

\section{Numerical study: Observables}
\label{sec:observables}

We are interested in studying two types of synchronizations: phase synchronization and frequency entrainment. The various statistical quantities that we measure in our study are as follows:

\subsection{Order parameter}
To measure frequency entrainment, we introduce the Edwards-Anderson order parameter, defined as~\cite{chowdhury2010synchronization}
\begin{align}
	r_{EA} e^{{\rm i} \psi_{EA}} \equiv \lim_{T \to \infty} \frac{1}{N}  \sum_{j=1}^{N} e^{{\rm i}
		\left[\theta_{j}(t_0 +T) - \theta_{j}(t_0) \right]},
	\label{eq:EA_op_definition}
\end{align}
where $t_0$ is the time larger than the initial transient time so that the dynamics settles down into a stationary state. The quantity $r_{EA}$ ($0 \leq r_{EA} \leq 1$ ) measures the amount of entrainment present in the system; $r_{EA}=1$ corresponds to a fully entrained phase, whereas $r_{EA} = 0$ corresponds to an unentrained phase. Note that by `entrainment', we mean the oscillators' stationary-state long-time average of the frequencies to be the same but not necessarily their instantaneous frequencies. 

To study the phase synchronization, let us also introduce the Kuramoto synchronization order parameter~\cite{Kuramotobook,Strogatz2000}
\begin{align}
	r(t) e^{{\rm i} \psi(t)} \equiv \frac{1}{N}  \sum_{j=1}^{N} e^{{\rm i}
		\theta_{j}(t)},
	\label{eq:Kura_op_definition}
\end{align}
where the quantity $r$ ($0\le r \le 1$) measures the amount of global phase synchrony present in the system at a given instant in time $t$, while $\psi \in [0,2\pi)$ measures the average phase at that instant~\cite{Strogatz2000}.

\subsection{Dynamic fluctuations}
We study the behavior of the stationary-state fluctuations of the frequency order parameter by measuring the quantity
\begin{eqnarray}
\chi = N \left[ \langle r_{EA}^2 \rangle -  \langle r_{EA} \rangle^2 \right].
\label{eq:def_fluc_EA}
\end{eqnarray}
In our simulations, to compute the frequency order parameter for a given realization of the dynamics, we first evolve the dynamics until it reaches a stationary state, signaled by a stationary distribution of the phase order parameter. In the stationary state, we choose time intervals of varying lengths  $T_n= T+ n \Delta T$, with $T=500$, $\Delta T = 10$, and $n$ running from $0, 1, 2, \cdots, 199$, to construct a distribution of the EA order parameter, the mean of which yields the time-averaged EA order parameter. The quantity thus computed is further averaged over many such realizations of the network and natural frequencies of the oscillators (sample average). Here, $\langle  \cdot \rangle$ and $\left[ \cdot \right]$ represent the time average in the stationary state and sample averages, respectively. This quantity is equivalent to the susceptibility in statistical systems. We note that in computing the time average, it is advisable to choose non-overlapping time intervals when constructing the distribution in order to eliminate any statistical correlations in the data. Clearly, this approach is computationally very expensive. However, once averaged over a long total time, we have observed that both of these schemes yield qualitatively similar behavior. 

A similar quantity for the stationary-state fluctuations of the phase order parameter is measured as follows
\begin{eqnarray}
\chi = N \left[ \langle r^2 \rangle -  \langle r \rangle^2 \right].
\label{eq:def_fluc_Kura}
\end{eqnarray}
Here, to compute the time-average value for a given realization of the dynamics, unlike in the previous case, one records only the values of phase order parameter at various time instants in the stationary state. 

\subsection{Binder Cumulant}
To understand the existence as well as the nature of a phase transition, we consider another useful thermodynamic quantity, namely, the fourth-order Binder cumulant, defined on system of size $N$ by \cite{binder1981finite, binder-book}
\begin{eqnarray}
U = 1- \left[\frac{\langle r^4 \rangle}{3\langle r^2 \rangle^{2}}\right],
\label{eq:def_Binder}
\end{eqnarray}
where $\langle \cdot \rangle$ and $[\cdot]$ represent the time average in the stationary state and the sample average, respectively. We replace $r$ in Eq.~($\ref{eq:def_Binder}$) by $r_{EA}$ to study frequency entrainment dynamics.

Our analysis is based on the  finite-size scaling (FSS) hypothesis and we further assume that this scaling holds also for continuous transitions in a nonequilibrium system, in particular for a large network in the limit $N \to \infty$\,\cite{binder2005finite, binder-book, hong2007finite, um2014nature, hong2015finite, coletta2017finite}. Similar to an equilibrium system, we assume that, except at criticality, the two-point connected correlation between two oscillator phases at $i$ and $j$ to behave exponentially $\sim \exp{(-|i-j|/\xi)}$ in both the ordered (entrained/synchronized) and disordered (unentrained/unsynchronized) phases, where $|i-j|$ is the separation between the two oscillators as defined earlier, and $\xi$ is called the correlation length. A continuous phase transition is characterized by a divergence of correlation length $\xi$ at criticality, and finite otherwise.

Following the FSS hypothesis, for large but finite $N$, the correlation length $\xi \ll N$ remains limited in both ordered and disordered phases (away from criticality). Consequently, $U$ converges to the asymptotic value $2/3$ in the ordered phase and $1/3$ in the disordered phase. At criticality, the finite system size ($N$) will cut off long-distance correlations, and hence, one would expect finite-size rounding off of critical-point singularities, implying $\xi \sim N$. The system is now expected to remain close to another fixed-point value $U^{*}$, independent of $N$. So, a feature of a continuous transition is hinted at by the existence of a common intersection point of the curves for $U$ vs. the relevant coupling or noise strength for a network of various sizes $N$. The common intersection point corresponds to the critical parameter value of the transition at which fluctuation diverges in the thermodynamic limit and $\xi \to \infty$.

However, in practice, due to statistical uncertainties, instead of a common intersection point, the curves may cross each other within a range of parameter values. To estimate the ``true'' critical parameter, one then needs to study a large system size, perform sample averaging and take into account finite-size effects. However, in our present work, our aim is not to estimate the critical point, but to investigate the very existence of a transition, if any. Data of numerical results reported in Sec~\ref{sec:Kura} and Sec.~\ref{sec:xy} are obtained by numerically integrating the corresponding deterministic (Kuramoto model) and stochastic ($XY$ model) governing dynamics employing the Runge-Kutta4 (RK4) and Euler-Maruyama algorithm respectively, with integration time step ${\rm d}t = 0.01$.


\begin{figure*}[ht!]
	\centering
	\includegraphics[scale=0.6]{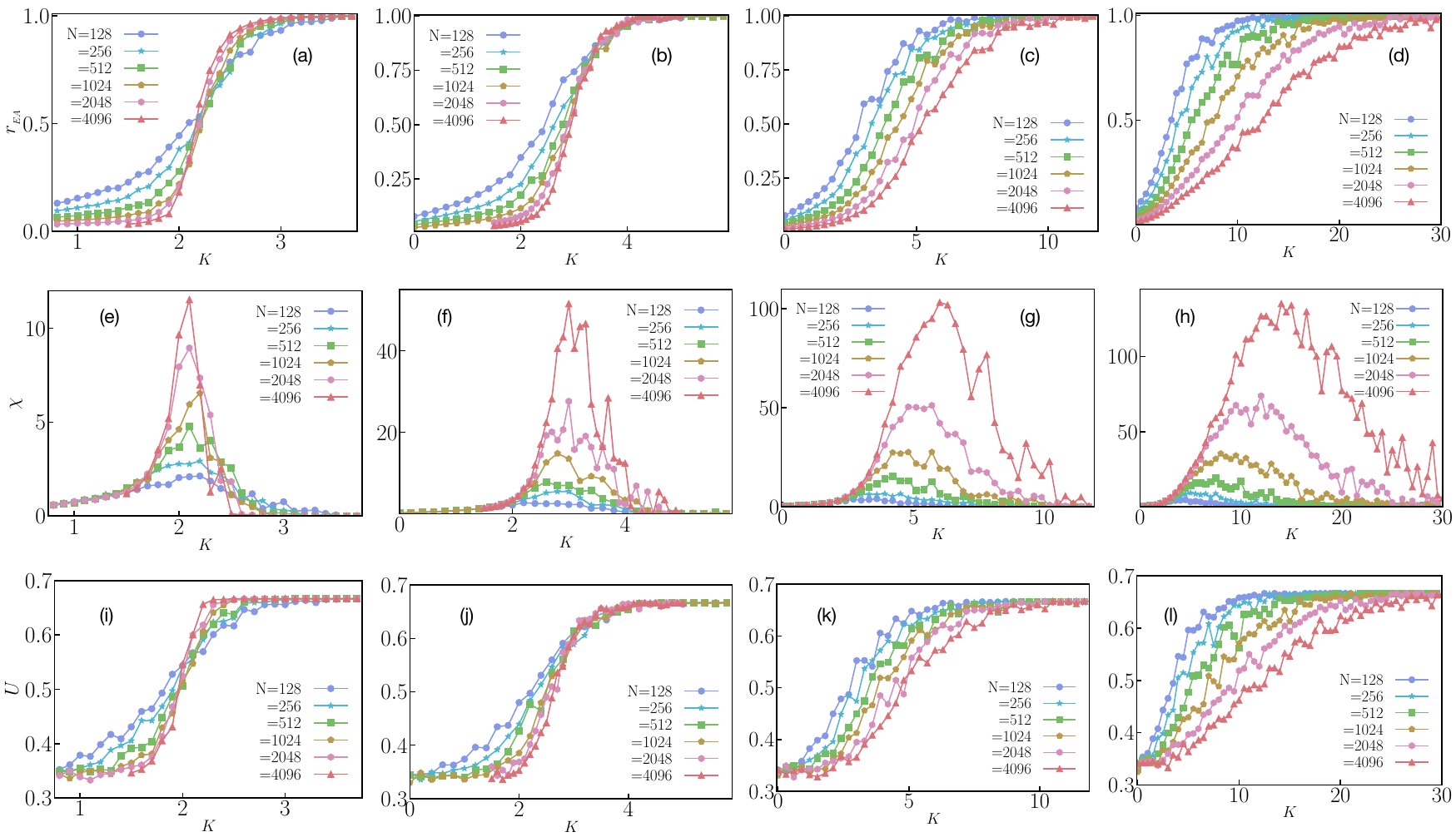} 
	\caption{Entrainment in Kuramoto dynamics: Stationary-state Edwards-Anderson order parameter $r_\text{EA}$ [panels (a-d)], dynamical fluctuations $\chi$  [panels (e-h)], and Binder cumulant $U$ [panels (i-l)] as a function of coupling strength $K$ for 4 different $\sigma$ values: $\sigma=0.2$ [panels (a), (e), (i)], $0.4$ [panels (b), (f), (j)], $0.6$ [panels (c), (g), (k)] and $0.8$ [panels (d), (h), (l)]. Data in each panel are obtained in the nonequilibrium stationary state by integrating the dynamics~(\ref{eq:eom0}) on networks of sizes $N=128,~256,~512,~1024, ~2048$, and $4096$ as indicated in the legend and averaged over $50$ different realizations of the network and intrinsic frequencies of the oscillators.}
	\label{fig:EA_entrainment}
\end{figure*}

\begin{figure*}[ht!]
	\centering
	\includegraphics[scale=0.6]{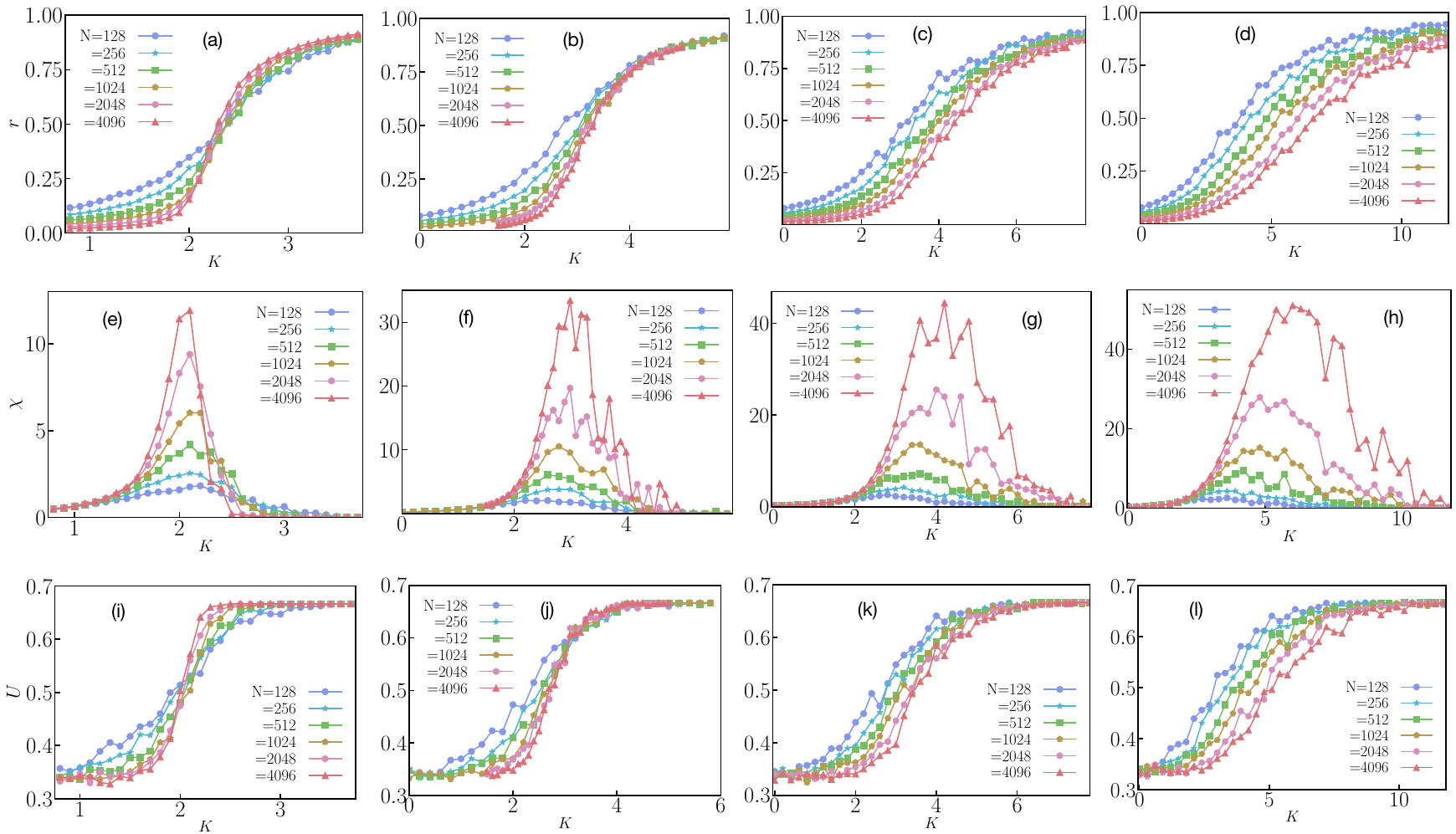}
	\caption{Phase synchronization in Kuramoto dynamics: Stationary-state Kuramoto order parameter [panels (a-d)], dynamical fluctuations $\chi$  [panels (e-h)], and Binder cumulant $U$ [panels (i-l)] as a function of coupling strength $K$ for four different $\sigma$-values, namely, $\sigma=0.2$ [panels (a), (e), (i)], $0.4$ [panels (b), (f), (j)], $0.5$ [panels (c), (g), (k)] and $0.6$ [panels (d), (h), (l)]. Data in each panel are obtained in the nonequilibrium stationary state by integrating the dynamics~(\ref{eq:eom0}) on networks of sizes $N=128,~256,~512,~1024, ~2048$, and $4096$ as indicated in the legend and averaged over $50$ different realizations of the network and intrinsic frequencies of the oscillators.}
	\label{fig:phase_sync_Kura}
\end{figure*}

\section{Results: Universality of phase and entrainment transition}
\label{sec:Kura}
\subsection{Frequency entrainment transition}
With the observables mentioned above, we now proceed to study the entrainment dynamics in the Kuramoto model on networks of various spectral dimensions. Figure~(\ref{fig:EA_entrainment}) shows the behavior of the stationary-state EA order parameter $r_{EA}$ [panels (a-d)], dynamical fluctuations $\chi$  [panels (e-h)], and Binder cumulant $U$ [panels (i-l)] as a function of coupling strength $K$ for various $\sigma$ values on networks of various sizes $N$.

For all $\sigma$ values [panels (a-d)], the EA order parameter increases from small values $\mathcal{O} (1/\sqrt{N})$ to large values $\mathcal{O}(1)$. However, to investigate whether these unambiguously correspond to an entrainment transition in the thermodynamic limit, we study the behavior of dynamical fluctuations [panels (e-h)] and Binder cumulant $U$ [panels (i-l)]. The divergence in $\chi$ at criticality in the thermodynamic limit is reflected in finite systems as a peak in the fluctuation curve, which is rounded at finite size. We further expect $\chi$ to be $\mathcal{O} (1)$ both in the ordered and disordered phase, in a region away from criticality. We call the parameter value at which the peak occurs a pseudo-critical point. A few comments on the Fig.~\ref{fig:EA_entrainment}(e-h) are in order: first, the peak height of $\chi$ increases with $N$; second, the fluctuation curves look asymmetric around their maxima, and thus they are expected to yield two different values of the critical exponent corresponding to either side of the transition. This may be due to the heterogeneity present in our system: in our model, each oscillator experiences a different field due to its own intrinsic frequency and different degree. Thus, the usual renormalization group argument for equal exponents above and below the transition based on a few relevant variables may not work in such a case. Third, the fluctuation curves around the peak in the region of high $K$ values become steeper as $N$ increases. All these observations hint towards the existence of a transition. 

At this point, a close scrutiny shows something very interesting. One expects, in general, the fluctuation curves around the peak in the region of high $K$ value to be steeper as $N$ increases, which implies diverging fluctuations only at the critical point and zero fluctuations away from criticality in the thermodynamic limit. This is indeed observed in panels (e-f), which correspond to $\sigma=0.2$ and $0.4$, or equivalently, $d_s = 10$ and $5$. These bear a clear signature of an entrainment transition in the limit $N \to \infty$, as expected. However, for $\sigma \geq 0.6$ or $d_s \lesssim 3.33$ the behavior changes, see panels (g-h) for $\sigma=0.6$ and $0.8$, or equivalently, $d_s \approx 3.3$ and $2.5$. The curves broaden and imply large fluctuations that increase with $N$ even for very large $K$ away from pseudo-criticality; this precludes the existence of an ordered/entrained phase at high $K$ in the thermodynamic limit. This shows that there is no entrained phase, and hence no entrainment-unentrainment transition for $\sigma \geq 0.6$, or $d_s \lesssim 3.33 $.

To confirm this observation, we further look at the behavior of the Binder cumulant $U$ [see Fig.~\ref{fig:EA_entrainment}(i-l)] as a function of coupling strength $K$ for the same $\sigma$ values. As observed in panel (i), the curves for various $N$ intersect within a certain range of $K$; also in panel (j), the behavior of $U$ for the two largest networks $N=2048$ and $4096$ shows an intersection in $U$ for large $K$ that indicates an entrainment transition. Qualitatively different behavior is seen in panels (k) and (l), i.e.~for $d_s \lesssim 3.33$, where the curves for different $N$ do not intersect. This validates our observation, as obtained from the study of dynamic fluctuations.

We note that linear theory predicts the lower critical dimension of the entrainment transition as $d_s =2$, given that the correlation does not diverge. Here, we observe the absence of entrainment even in dimensions  $(d_s \lesssim 3.33)$ higher than the critical one. We believe this happens because of an enhanced fluctuations arising from the nontrivial interplay between the `quenched' network disorder and the quenched frequency disorder, besides the nonlinearity in the dynamics, suppressing the entrained phase predicted from linear theory in this region.


\subsection{Phase synchronization transition}
In this section, we study the phase synchronization dynamics of the Kuramoto model on our network of various spectral dimensions and investigate the critical dimension to observe the phase transition. Figure~\ref{fig:phase_sync_Kura} shows the behavior of the stationary-state Kuramoto order parameter $r$ [panels (a-d)], dynamical fluctuations $\chi$  [panels (e-h)], and Binder cumulant $U$ [panels (i-l)] as a function of coupling strength $K$ for various $\sigma$ values on networks of various sizes $N$.

Similar to the entrainment case, for all $\sigma$ values [panels (a-d)], the Kuramoto order parameter also increases from small values $\mathcal{O} (1/\sqrt{N})$ to large values $\mathcal{O}(1)$. The behavior of the dynamic fluctuations of the Kuramoto order parameter for $\sigma=0.2$ and $0.4$, or, equivalently, $d_s=10$ and $5$ [panels (e-f)] shows the existence of a phase synchronization transition in the limit $N \to \infty$. The case for $\sigma=0.5$, which corresponds to $d_s=4$ [panel (g)], is the marginal case as predicted from the linear theory; furthermore, panel (h) shows the behavior at $\sigma=0.6$, or $d_s \approx 3.33$. In the latter two cases, the peaks of the fluctuation curves start to broaden for larger system size $N$ in the region of high $K$, and thus imply large fluctuations in the thermodynamic limit.  Thus, in the region $\sigma \geq 0.5$ ($d_s \leq 4$), the broadening of the peak supports having no synchronized phase at any high $K$, and thus no phase synchronization-desynchronization transition.

This observation is further confirmed from the behavior of the Binder cumulant $U$ [Fig.~\ref{fig:phase_sync_Kura} (i-l)] as a function of coupling strength $K$: the curves for various sizes $N$ cross in panels (i) and (j), which corresponds to $d_s=10$ and $5$, respectively. However, having no common intersection point among the various curves of $U$ in panels (k) and (l) bears a clear signature of no phase synchronization transition for $\sigma\geq 0.5$, i.e.~$d_s \leq 4$. 

Our observation of the lower critical dimension of the phase synchronization transition is thus consistent with the prediction of linear theory in that the transition occurs only in $d_s > 4$. One may thus conclude that the network disorder does not affect the critical dynamics of the phase synchronization transition.


\begin{figure*}[ht!]
	\centering
	\includegraphics[scale=0.6]{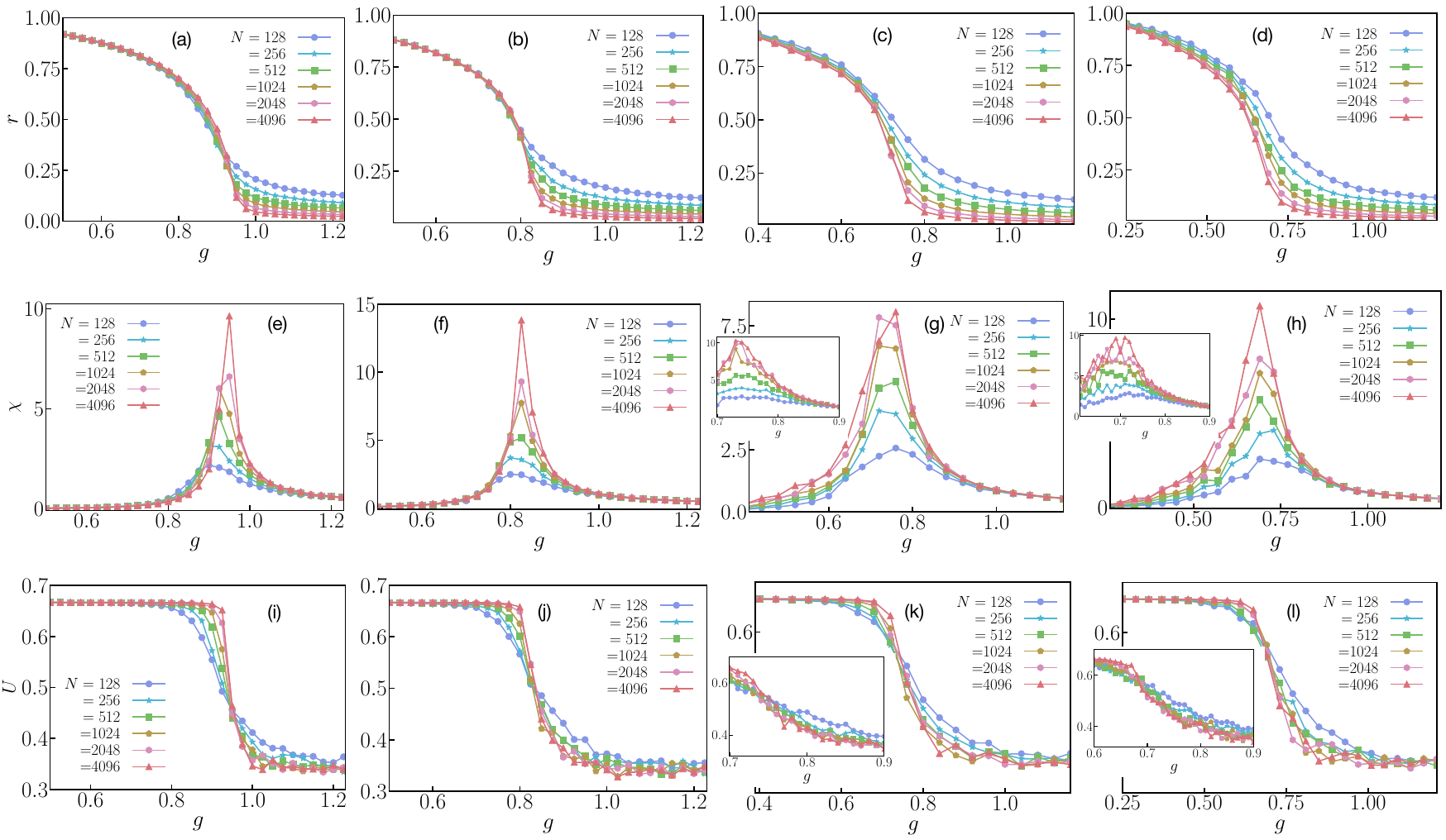}
	\caption{XY model: Stationary-state phase order parameter $r$ [panels (a-d)], dynamical fluctuations $\chi$  [panels (e-h)], and Binder cumulant $U$ [panels (i-l)] as a function of noise strength $g=\sqrt{2T/K}$ for four different $\sigma$ values, namely, $\sigma=0.1$ [panels (a), (e), (i)], $0.5$ [panels (b), (f), (j)], $0.7$ [panels (c), (g), (k)] and $0.8$ [panels (d), (h), (l)]. Data in each panel are obtained in the nonequilibrium stationary state by integrating the dynamics~(\ref{eq:eom_GWN2}) on networks of sizes $N=128,~256,~512,~1024$, and $2048$ as indicated in the legend and averaged over $50$ different realizations of the network and noise.}
	\label{fig:phase_sync_XY}
\end{figure*}

\section{XY model limit: Universality of phase and BKT transitions}

\label{sec:xy}
So far, our study focuses on the role of network disorder on the emergent dynamics of a nonequilibrium system. At this point, we feel it worthwhile to explore its effect on the equilibrium dynamics. To this end, we consider the $XY$ model limit of the dynamics~(\ref{eq:eom0}). Note that network disorder in our study eventually renders the XY dynamics of a nonequilibrium nature.  As already predicted by the linear theory, the lower critical dimension for a phase transition in the $XY$ model on a general network is $d_s^{l} = 2$. In this section, we aim to verify the universality of this phenomenon. The $XY$ model is obtained from the usual Kuramoto model when the intrinsic frequencies of the oscillators are chosen to be identical, and moreover set to zero, and then subject to Gaussian white noise~\cite{sarkar2020noise,sarkar2021synchronization}. The phase evolution equation (\ref{eq:eom0}), in this limit, now reads as
\begin{align}
	\frac{{\rm d} \theta_{i}}{{\rm d}t}= \frac{K}{\kappa_i}\sum_{j=1}^N  a_{ij} \sin(\theta_{j} -\theta_{i}) + \eta_i (t),
	\label{eq:eom_GWN0}
\end{align}
where the term $\eta_{i}(t)$ is a Gaussian white noise characterized by
\begin{align}
	\langle \eta_{i}(t) \rangle = 0 ~\text{and}~ \langle \eta_{i}(t)\eta_{j}({t'}) \rangle = 2T\delta_{ij} \delta(t - {t}^{\prime}).
\end{align}
Here $\langle \cdot \rangle$ denotes averaging over noise realizations, and $T$ is the noise strength proportional to the temperature of the system.  One may view the system of identical oscillators to be in contact with a heat bath which is at a temperature $T$. To show the explicit $T$ dependence, one could write Eq.~(\ref{eq:eom_GWN0}) as
\begin{align}
	\frac{{\rm d} \theta_{i}}{{\rm d}t}= \frac{K}{\kappa_i}\sum_{j=1}^N  a_{ij} \sin(\theta_{j} -\theta_{i}) + \sqrt{2T} \zeta_{i}(t),
	\label{eq:eom_GWN1}
\end{align}
with
\begin{align}
	\langle \zeta_{i}(t) \rangle = 0 ~\text{and}~ \langle \zeta_{i}(t)\zeta_{j}({t'}) \rangle = \delta_{ij} \delta(t - {t}^{\prime}).
\end{align}
For $K \neq 0$ the equation of motion (\ref{eq:eom_GWN1}) can be brought into dimensionless form by the transformation
${t} \to Kt$, ${g} \to  \sqrt{2T/K}$ and ${\zeta}_{i}({t}) \to {\zeta}_{i}(t)/K$,
\begin{align}
\frac{{\rm d} \theta_{i}}{{\rm d} t} =   \frac{1}{\kappa_i}\sum_{j=1}^N  a_{ij} \sin(\theta_{j} -\theta_{i})+ {g} {\zeta}_{i}({t}).
\label{eq:eom_GWN2}
\end{align}
The dynamics relaxes at long times to a nonequilibrium stationary state. The Kuramoto phase-order parameter $r$ is now equivalent to the magnetization in a spin model of statistical physics. Note that, for convenience, all numerical results in this section are presented as a function of the reduced noise strength $g$ that depends on the temperature $T$ as $g = \sqrt{2T/K} $. However, we use both the terms `reduced noise strength' and `temperature' interchangeably throughout the text.

\begin{figure}
	\centering
	\includegraphics[scale=0.35]{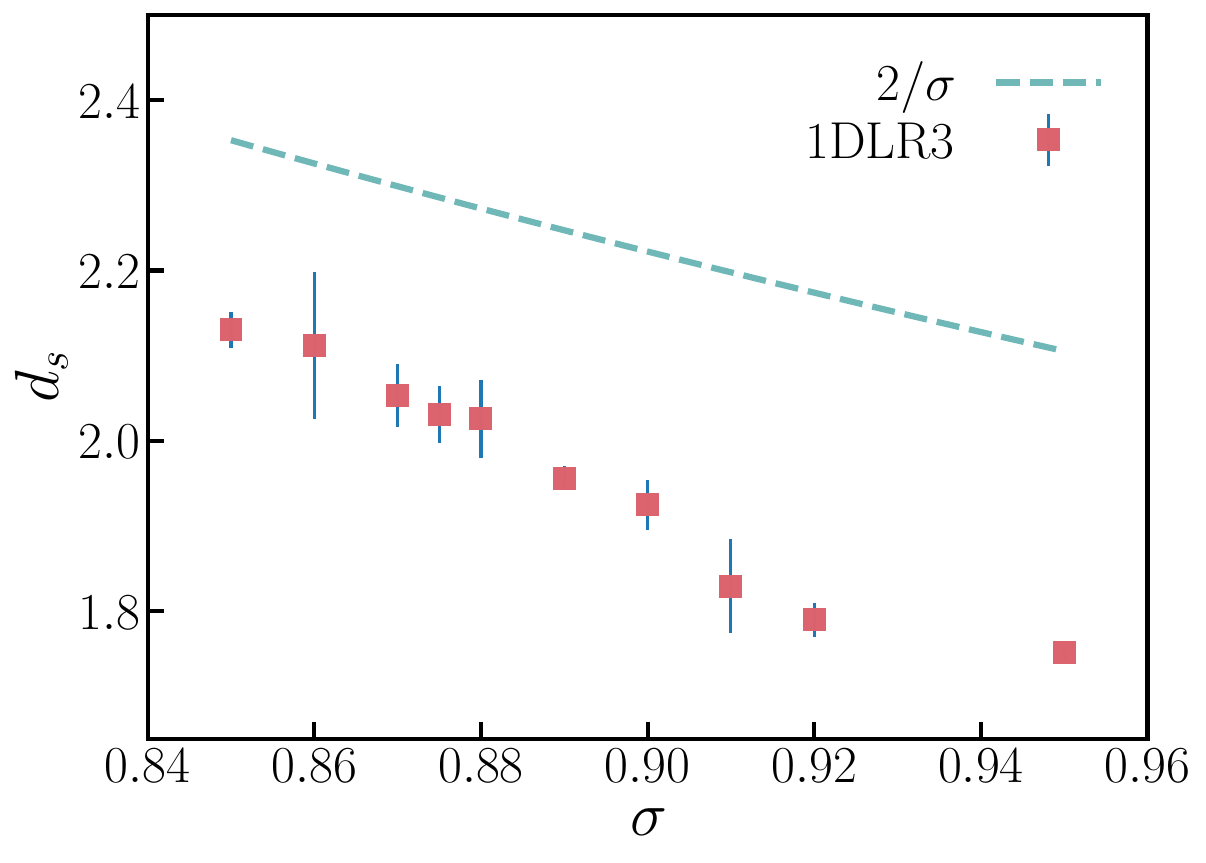}
	\caption{Spectral dimension $d_s$ as a function of $\sigma$ of the $1DLR3$ network as obtained by the finite-size scaling of the graph Laplacian spectrum. The dashed blue line represents the analytical expectation for a long-range weighted lattice.}
	\label{fig:ds_vs_sigma}
\end{figure}

\begin{figure*}[ht!]
	\centering
	\includegraphics[scale=0.6]{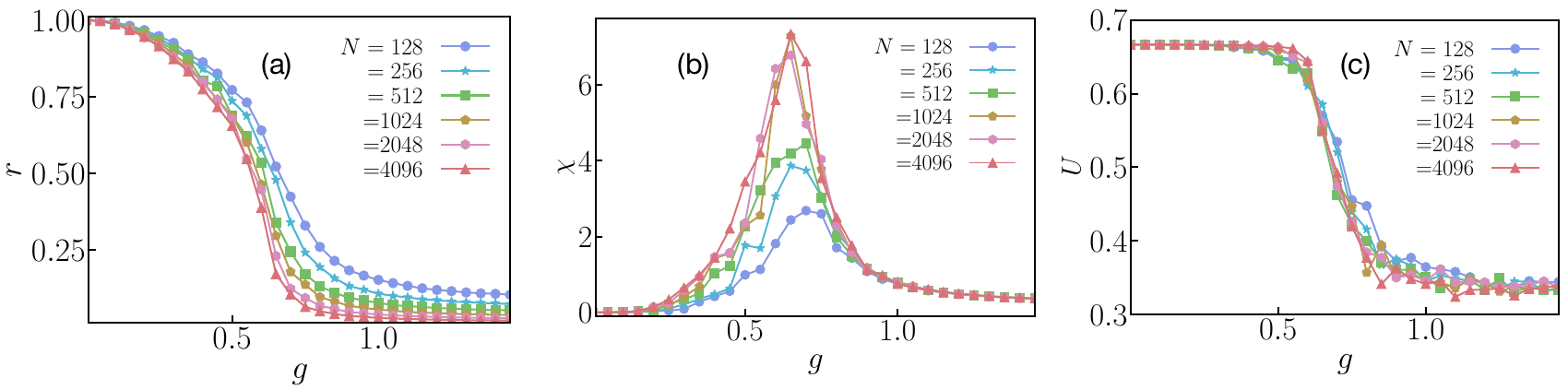}
	\caption{BKT transition in XY model: Stationary-state phase order parameter $r$[panels (a)], dynamical fluctuations $\chi$  [panels (b)], and Binder cumulant $U$ [panels (c)] as a function of reduced noise strength $g$ for $\sigma=0.875$ that corresponds to $d_s \approx 2.0$. Data in each panel are obtained in the nonequilibrium stationary state by integrating the dynamics~(\ref{eq:eom_GWN2}) on networks of sizes $N=128,~256,~512,~1024, ~2048$, and $4096$ as indicated in the legend and averaged over $50$ different realizations of the network and noise.}
	\label{fig:LR3_no_BKT}
\end{figure*}

\subsection{Phase transition}

To study the phase transition, similar analysis of the various statistical quantities as in the previous sections is pursued, and is shown in Figure~\ref{fig:phase_sync_XY}.
The stationary-state phase order parameter [panels (a-d)] shows a transition as temperature increases in $r$-value from $\mathcal{O}(1)$ to $\mathcal{O}(1/\sqrt{N})$ on finite networks of size $N$.
The behavior of the fluctuations, that the peak height increases with $N$, and especially that the fluctuation curves become steeper near pseudo-criticality in the region of the ordered phase, shows the existence of a phase transition for $\sigma=0.1$ and $0.5$, i.e., $d_s=20$ and $4$ [panels(e-f)].
However, as $\sigma \geq 0.7$, e.g., in panels (g,h) corresponding to $\sigma=0.7$ and $\sigma=0.8$, or equivalently $d_s\approx 3$ and $2.5$, the maximum fluctuations appear to saturate with system size $N$ (see insets). Also, the curves broaden with increasing $N$ near their maxima in the region of the ordered phase, probably caused by strong finite size effects that hinder the realization of the ordered phase, whose existence has been proven in the thermodynamic limit\,\cite{vezzani1999inverse}. Further, we show the behavior of the Binder cumulant $U$ in Fig.~\ref{fig:phase_sync_XY}(i-l) for the same $\sigma$ values. The curves of $U$ vs.\ $g$ for various $N$ seem to intersect clearly for $\sigma= 0.1$ and $0.5$ [panels (i) and (j)], whereas for $\sigma= 0.7$ and $0.8$ [panels (k) and (l)] their $N$-dependent behavior in the region of the ordered phase does not show a clear intersection, consistent with our previous observation from dynamic fluctuations. Thus, in the region $\sigma \gtrsim 0.7$, i.e.~$d_s \lesssim 3$, we do not obtain a clear signature of a phase transition. The missing evidences for the ordered phase in~$d_s \lesssim 3$ ought to be the result of the strong finite-size effects generated by disorder as the model has been theoretically shown to magnetize in the thermodynamic limit for all $d_{s}>2$\,\cite{vezzani1999inverse}.

\subsection{Fate of the \textit{BKT} transition}
So far, our study focused on the existence of a conventional phase transition on a network. This section is devoted to studying the Berezinskii-Kosterlitz-Thouless ($BKT$) transition on networks~\cite{kosterlitz1973ordering,kosterlitz1974critical}. As stated earlier, the Mermin-Wagner theorem states for an equilibrium system that a continuous symmetry cannot be broken spontaneously at any finite temperatures in spatial dimensions two or lower~\cite{mermin1966absence}. The 2D $XY$ model exhibits an unconventional phase transition, the $BKT$ transition, between a low-temperature quasi-ordered phase and high-temperature disordered phase \cite{kosterlitz1973ordering,kosterlitz1974critical}. The quasi long-range ordered phase is characterized by an algebraic decay of correlations (infinite correlation length and diverging susceptibility), whereas the decay is exponential in the disordered phase. A generalization of the Mermin-Wagner theorem is as follows: No spontaneous breaking of continuous symmetry is possible on a ``recursive on the average'' graph, i.e.~on a graph with ``average spectral dimension'' $d_s \leq 2$. The continuous symmetry models always have a broken symmetry phase at a finite temperature on the ``transient on the average'' graph ($d_s > 2$). Note that this is consistent with our theoretical prediction for the linear model.

Motivated by this, we ask if the $XY$ dynamics on 1DLR3 network with $d_s=2$ shows a $BKT$ transition. In order to realize $d_s = 2$ with our network, we numerically compute $d_s$ from finite-size scaling of low-lying eigenvalues of the graph Laplacian for various $\sigma$, which yields $\sigma=0.875$ corresponding to $d_s \approx 2.03$; see Fig.~{\ref{fig:ds_vs_sigma}}. The behavior of various statistical quantities measured in the stationary state at this parameter value is shown in Fig.~\ref{fig:LR3_no_BKT}.

Now, for a system exhibiting a $BKT$ transition at critical temperature $T_{BKT}$, in the region $T \leq T_{BKT}$, fluctuations diverge, and consequently the correlation length $\xi(N \to \infty)$ is infinite. Thus, for large but finite $N$, one would expect the fluctuations to scale with $N$ at and below $T_{BKT}$. Also, the curves of $U$ for various $N$ are expected to stay close to a fixed point $U^{*}$ in this region. However, in practical application, due to the statistical uncertainties, it is observed that $T_{BKT}$ is very close to the point where the curves of $U$ for various $N$ begin to separate from the low-$T$ asymptotic value \cite{wysin2005extinction}.

Figure~\ref{fig:LR3_no_BKT}(a) shows the behavior of the stationary order parameter $r$ as a function of noise strength $g$, which crosses over from low-temperature high $r$-value to a high temperature low $r$-value, typical of finite-size rounding of a sharp transition. However, as shown in panel (b), the fluctuations in the low-temperature region do not seem to increase with $N$ (even the peak heights seem to get saturated) for high values of $N$, showing no divergence in this region in the thermodynamic limit. This is further confirmed by the behavior of the Binder cumulant $U$ [panel (c)]: the curves of $U$ for various $N$ do not seem to stay collapsed and then begin to separate from a low-$g$ asymptotic value, nor even do they intersect clearly. This evidence speaks against a BKT transition in disordered graphs with $d_s=2$; instead, its behavior is similar to that observed for $\sigma=0.7$ and $0.8$ ($d_s>2$).

Based on these observations, we anticipate that the network disorder plays a crucial role in $0.6 < \sigma < 0.875$, or approximately, in spectral dimension $2 \leq d_s \lesssim 3$. The structural heterogeneity due to the presence of the long-range links acts as a source of quenched disorder. The lower critical dimension of the phase synchronization transition is $d_s=4$, and in such high dimension the disorder fluctuation decreases with system sizes very fast, and the system becomes asymptotically homogeneous at large length scales, suppressing the fluctuations. Thus, the critical behavior of the phase synchronization transition in such disordered systems remains unaffected and is identical to that of the clean system. In contrast, for both the entrainment dynamics of the Kuramoto model and the phase transition of the $XY$ model, the lower critical dimension is $d_s = 2$. In $2 \leq d_s \lesssim 3$, we believe that such a weak disorder makes the finite-size effects very strong, introducing enhanced fluctuations that vanish extremely slowly in the thermodynamic limit. Thus, the system remains inhomogeneous even at large length scales (for the system sizes we considered), making it extremely difficult to probe the existence of the ordered phase predicted by the linear theory and also by rigorous mathematical arguments\,\cite{vezzani1999inverse}. However, a complete confirmation of this calls for an independent study.

\section{Conclusion}
\label{sec:conclusion}
In this work, we have systematically investigated the role of spectral dimension $d_s$ as a control parameter in determining the universality of phase transitions on a network. As a network, we employ a 1DLR3 graph on top of which we study two paradigmatic models: One is the non-equilibrium dynamics of the Kuramoto model of synchronization, and the other one is the equilibrium dynamics of the classical $XY$ model. To summarize our findings,
\begin{enumerate}
    \item We have developed for a given dynamics occurring on the network, under linear approximation, a general relationship between stationary-state properties of the dynamics occurring on the network and the underlying network structures in terms of the density of eigenvalues of the network Laplacian. Our method is general in that it applies to both deterministic and stochastic dynamics on the network so long as it has a unique stationary state.

    \item The linear theory predicts the lower critical spectral dimension for the entrainment and synchronization transition in the Kuramoto model as $d_s=2$ and $d_s=4$, respectively, whereas, for the phase transition in the $XY$ model, it yields $d_s=2$. 

    \item Our detailed numerical investigation agrees well with the theoretical prediction of phase synchronization transition in the Kuramoto model. However, it does not yield a clear signature of entrainment transition in the Kuramoto model and phase transition in the $XY$ model in $2 \leq d_s \lesssim 3$. 

\end{enumerate}

Let us now briefly compare our results with that of the clean counterparts of the dynamics under study, i.e., dynamics on a $d$-dimensional periodic lattice with long-range (LR) power-law interactions $\sim |i-j|^{-(d+\sigma)}$, $|i-j|$ being the separation between two sites on the lattice, as defined earlier. The Kuramoto model with LR interactions on a 1D lattice ($d=1$) has been numerically studied previously\,\cite{rogers1996phase, marodi2002synchronization, chowdhury2010synchronization}. Numerical simulations in Ref.\,\cite{rogers1996phase} demonstrate that an entrainment transition is possible only when $\sigma  \leq \sigma_c = 1$, which implies a lower critical spectral dimension for an entrainment transition of $d_s=2$. On the other hand, Ref.\,\cite{kunz1976first} provides a rigorous mathematical proof on the existence of stable spontaneous magnetization at finite temperature in the 2D XY model ($d=2$) with long-range (LR) interactions for $\sigma < 2$, suggesting a lower critical spectral dimension for phase transition of $d_s = 2$. Thus, both clean systems, the LR Kuramoto and LR XY models, indeed exhibit an entrainment and a continuous transition, respectively, above $d_s = 2$, thereby confirming our expectations from linear theory. We emphasize that the only difference between our work and Refs.\,\cite{kunz1976first, rogers1996phase} is the presence of network disorder, which seems to have a crucial impact on the emergent phenomena.

We thus anticipate that the quenched network disorder arising from the structural heterogeneity is harmless in spectral dimension $d_s >3$ and the critical behavior in such disordered systems is identical to that of the clean system. However, for $2 \leq d_s \lesssim 3$ the disorder average introduces finite-size scaling contributions which vanish extremely slowly (possibly logarithmically) in the thermodynamic limit, making it extremely hard to probe the existence of the ordered phase predicted by the linear theory as well as by rigorous mathematical arguments\,\cite{vezzani1999inverse}.
Our next task would be to develop a rigorous theory, such as a field-theoretic approach using a functional renormalization group, to establish our findings. This further includes computing the critical exponents of the associated transitions and determining the universality classes. Moreover, given the role of network disorder near $d_s=2$, an immediate question arises: How can one realize a $BKT$ transition in the dynamics of $XY$ model on such a disordered system? A complete confirmation of such a nontrivial behavior and an answer to the imposed query require an independent study. Investigation in this direction is going on and will be reported elsewhere.

\begin{acknowledgments}
We acknowledge fruitful discussions with Giacomo Gori. This project is supported by the Deutsche Forschungsgemeinschaft (DFG, German Research Foundation) under Germany’s Excellence Strategy EXC 2181/1-390900948 (the Heidelberg STRUCTURES Excellence Cluster). M.S. also acknowledges support by the state of Baden-Württemberg through bwHPC cluster.
\end{acknowledgments}

\appendix
\section{Linear theory: Derivation of  $\left \langle  \theta_{\lambda}^{L} (t) \theta_{\lambda}^{R} (t) \right \rangle $}
\label{app:Linear_theory}
Here we present a detailed derivation of various quantities, evolution equations in the new variables, for the linearized dynamics of both the Kuramoto and $XY$ models. To analyze the dynamics~(\ref{eq:eom_linear}), as mentioned in the main text, we work in the eigenbasis of the asymmetric Laplacian $ \mathbf {\mathcal L}$ . If $ \ket {v_{m}^{R}}$ and $\bra {v_{m}^{L}}$ be the right and left eigenvectors respectively corresponding to an eigenvalue $\lambda_m$, we can represent a state, given by the phases of the oscillators, $\ket \theta = (\theta_1, \theta_2, \cdots, \theta_L)^{\intercal}$ in an eigenbasis as follows:
\begin{align}
\ket {\theta }= \sum_{m =1}^{N} \qprod{v_{m}^{L}}{ \theta} \ket{v_{m}^{R}} = \sum_{m=1}^{N} \theta_{\lambda_{m}}^{R}  \ket{v_{m}^{R}},\\
\bra {\theta }= \sum_{m =1}^{N} \qprod{ \theta} {v_{m}^{R}}  \bra{v_{m}^{L}} = \sum_{m=1}^{N} \theta_{\lambda_{m}}^{L}  \bra{v_{m}^{L}}.
\label{eq:theta_eigen_App}
 \end{align}
Similarly, a given realization of the natural frequencies and noise can also be represented by
\begin{align}
\ket {\omega} = \sum_{m=1}^{N} \omega_{\lambda_{m}}^{R}  \ket{v_{m}^{R}},~ \ket {\eta} = \sum_{m=1}^{N} \eta_{\lambda_{m}}^{R}  \ket{v_{m}^{R}},\\
\bra {\omega } = \sum_{m=1}^{N} \omega_{\lambda_{m}}^{L}  \bra{v_{m}^{L}},~\bra {\eta } = \sum_{m=1}^{N} \eta_{\lambda_{m}}^{L}  \bra{v_{m}^{L}}.
\label{eq:omega_eta_eigen_App}
 \end{align}

\subsection {Derivation of $ \left \langle  \theta_{\lambda}^{L} (t) \theta_{\lambda}^{R} (t) \right \rangle$  for the Kuramoto model}
We obtain from the linearized Kuramoto dynamics~Eq.~(\ref{eq:eom_linear}), when projected along the eigenbasis, 
\begin{align}
\frac{{\rm d} \theta_{\lambda_m}^{R}}{{\rm d}t} &= -  K  \lambda_{m} \theta_{\lambda_m}^{R}  +  \omega_{\lambda_{m}}^{R} \nonumber ,\\
\frac{{\rm d} \theta_{\lambda_m}^{L}}{{\rm d}t} &= -  K  \lambda_{m} \theta_{\lambda_m}^{L}  +  \omega_{\lambda_{m}}^{L} , ~~m = 1, 2, 3, \cdots, N.
\label{eq:eom_Kura_eigen}
\end{align}
whereas the frequency $\omega_{\lambda_{m}}^{L,R} $ satisfies
\begin{align}
\langle  \omega_{\lambda_{m}}^{L,R}   \rangle &= 0, \nonumber \\
\langle  \omega_{\lambda_{m}}^{L} \omega_{\lambda_{m'}}^{R }  \rangle &=  \delta_{\lambda_m, \lambda_{m'}}.
\label{eq:omega_corr_eigen}
\end{align}
Note that the frequencies $\omega_{\lambda_{m}}^{L,R}$ for a given $\lambda_m$ are $\delta$-correlated. The dynamics gets decoupled in the eigen basis and and we obtain $N$ independent first order stochastic differential equations. From now on, we use $\omega_{\lambda}^{L,R}$ instead of $\omega_{\lambda_{m}}^{L,R} $ for notational convenience. Solution to Eq.~(\ref{eq:eom_Kura_eigen}) is given by, for $\lambda >0 $,
\begin{align}
 \theta_{\lambda}^{L/R} (t) = \theta_{\lambda}^{L/R} (0)\, {\rm e}^{- K \lambda t} + \frac{\omega_{\lambda}^{L/R}}{K \lambda} \left( 1 - {\rm e}^{- K \lambda t}  \right).
 \label{eq:theta_lambda_soln_Kura}
\end{align}

Thus we have
 \begin{align}
& \left \langle  \theta_{\lambda}^{L} (t) \theta_{\lambda}^{R} (t) \right \rangle =  \theta_{\lambda}^{L} (0) \theta_{\lambda}^{R} (0) \, {\rm e}^{- 2 K \lambda t} \nonumber \\
 & + \frac{1}{K^2 \lambda^2}   \left( 1 - {\rm e}^{- K \lambda t}  \right)^2  \,\left \langle  \omega_{\lambda}^{L} \omega_{\lambda}^{R}  \right \rangle,\\
 &=    \theta_{\lambda}^{L} (0) \theta_{\lambda}^{R} (0) \, {\rm e}^{- 2 K \lambda t} + \frac{1}{K^2 \lambda^2}   \left( 1 - {\rm e}^{- K \lambda t}  \right)^2 .
\end{align}
At long times ($t \to \infty)$, it reduces to
 \begin{align}
  \left \langle  \theta_{\lambda}^{L} (t) \theta_{\lambda}^{R} (t) \right \rangle =   \frac{1}{K^2 \lambda^2}.
 \label{eq:theta_lambda_corr_Kura_app}
\end{align}
Equation~(\ref{eq:theta_lambda_corr_Kura_app}) is provided in the main text. Note that this result is already provided in Ref.~\cite{millan2019synchronization}. However, we give the derivation here in order to make our manuscript self-contained and we generalize this to the case of Gaussian white noise in the next section. 

\subsection {Derivation of $\left \langle  \theta_{\lambda}^{L} (t) \theta_{\lambda}^{R} (t) \right \rangle$ for the \textit{XY} model}
The governing dynamics (\ref{eq:eom_GWN0}), once projected along the eigenbasis, yields
\begin{align}
\frac{{\rm d} \theta_{\lambda_m}^{R}}{{\rm d}t} &= -  K  \lambda_{m} \theta_{\lambda_m}^{R}  +  \eta_{\lambda_{m}}^{R} \nonumber ,\\
\frac{{\rm d} \theta_{\lambda_m}^{L}}{{\rm d}t} &= -  K  \lambda_{m} \theta_{\lambda_m}^{L}  +  \eta_{\lambda_{m}}^{L} , ~~m = 1, 2, 3, \cdots, N.
\label{eq:eom_XY_eigen}
\end{align}
whereas, the noise $\eta_{\lambda_{m}}^{L,R} (t)$ can easily be shown to follow
\begin{align}
\langle  \eta_{\lambda_{m}}^{L,R} (t)  \rangle &= 0, \nonumber \\
\langle  \eta_{\lambda_{m}}^{L} (t) \eta_{\lambda_{m'}}^{R } (t') \rangle &= 2T \delta_{\lambda_m, \lambda_{m'}} \delta(t-t').
\label{eq:eta_corr_eigen}
\end{align}

Here, as well, the noise $\eta_{\lambda_{m}}^{L,R} (t)$ for a given $\lambda_m$ are $\delta$-correlated, and the dynamics get decoupled in the eigenbasis. Using the notation $\eta_{\lambda}^{L,R}$ instead of $\eta_{\lambda_{m}}^{L,R} $, Eq.~(\ref{eq:eom_XY_eigen}) yields a formal solution for $\lambda > 0$
 \begin{align}
 \theta_{\lambda}^{L/R} (t) = \theta_{\lambda}^{L/R} (0)\, {\rm e}^{- K \lambda t} + {\rm e}^{- K \lambda t}  \int_{0}^{t} {\rm d}t'\,\eta_{\lambda}^{L/R}(t')\, {\rm e}^{K \lambda t'} 
 \label{eq:theta_lambda_soln_XY}
\end{align}
Note that Eq.~(\ref{eq:theta_lambda_soln_XY}) implies for ensemble-averaged 
 \begin{align}
  \left \langle  \theta_{\lambda}^{L/R} (t) \right \rangle =   \theta_{\lambda}^{L/R} (0)\, {\rm e}^{- K \lambda t},
 \label{eq:theta_lambda_avg_XY_app}
\end{align}
which vanishes as $t \to \infty$.\\
One obtains for the quantity
 \begin{align}
& \left \langle  \theta_{\lambda}^{L} (t) \theta_{\lambda}^{R} (t) \right \rangle =  \theta_{\lambda}^{L} (0) \theta_{\lambda}^{R} (0) \, {\rm e}^{- 2 K \lambda t} \nonumber \\
 & + {\rm e}^{- 2 K \lambda t}  \int_{0}^{t} {\rm d}t' \int_{0}^{t} {\rm d}t''\,\left \langle  \eta_{\lambda}^{L} (t') \eta_{\lambda}^{R} (t'') \right \rangle\, {\rm e}^{K \lambda (t' + t'')},\nonumber \\
 &=   \theta_{\lambda}^{L} (0) \theta_{\lambda}^{R} (0) \, {\rm e}^{- 2 K \lambda t} + 2T {\rm e}^{- 2 K \lambda t}  \int_{0}^{t}  {\rm d}t' \, {\rm e}^{2K \lambda t' },\nonumber \\
 &=  \theta_{\lambda}^{L} (0) \theta_{\lambda}^{R} (0) \, {\rm e}^{- 2 K \lambda t} + \frac{T}{K \lambda} \left( 1 - {\rm e}^{- 2 K \lambda t}  \right),
\end{align}
which yields in the long time limit ($t \to \infty)$
 \begin{align}
  \left \langle  \theta_{\lambda}^{L} (t) \theta_{\lambda}^{R} (t) \right \rangle =   \frac{T}{K \lambda}.
 \label{eq:theta_lambda_corr_XY_app}
\end{align}
Equation~(\ref{eq:theta_lambda_corr_XY_app}) is provided in the main text.

\bibliography{References_mrinal}

\end{document}